\documentclass[conference]{IEEEtran}
\input epsf
\usepackage[utf8]{inputenc}

\usepackage{amsmath,amssymb,amsthm,graphics,amsfonts,graphicx,float}
\usepackage{lineno}
\usepackage{tikz}
\usepackage{ifthen}
\usepackage{ifpdf}
\usepackage{color}
\usepackage{enumerate}
\usepackage{enumitem}
\usepackage{caption}
\usepackage{subcaption}
\usetikzlibrary{arrows, decorations.pathmorphing, backgrounds, positioning, fit, petri, shapes}
\usepackage{algorithm}
\usepackage{algpseudocode}

\usepackage{placeins}
\usepackage{sidecap}
\usepackage{rotating}

\usepackage{array,multirow,makecell}
\setcellgapes{1pt}
\makegapedcells
\newcolumntype{R}[1]{>{\raggedleft\arraybackslash }b{#1}}
\newcolumntype{L}[1]{>{\raggedright\arraybackslash }b{#1}}
\newcolumntype{C}[1]{>{\centering\arraybackslash }b{#1}}

\usepackage{pdflscape}

\usepackage{fancyhdr} 
\begin{document}
	
	\title{\LARGE Energy-efficient workflow scheduling based on workflow structures under deadline and budget constraints in the cloud}
	
	\DeclareRobustCommand*{\IEEEauthorrefmark}[1]{%
		\raisebox{0pt}[0pt][0pt]{\textsuperscript{\footnotesize #1}}%
	}
	
	\author{\IEEEauthorblockN{J. E. {NDAMLABIN MBOULA}\IEEEauthorrefmark{1,}\IEEEauthorrefmark{2,}\IEEEauthorrefmark{3},
			V. C. KAMLA\IEEEauthorrefmark{1},
			M. H. HILMAN\IEEEauthorrefmark{4},  and
			C. {TAYOU DJAMEGNI}\IEEEauthorrefmark{5}}
		\IEEEauthorblockA{\IEEEauthorrefmark{1}Laboratory of Experimental Mathematics (LAMEX), University of Ngaoundere, Cameroon}
		\IEEEauthorblockA{\IEEEauthorrefmark{2}Inria Nancy – Grand Est, CAMUS Team, Villers-lès-Nancy, France}
		\IEEEauthorblockA{\IEEEauthorrefmark{3}ICube Laboratory, ICPS Team, Illkirch, France}
		\IEEEauthorblockA{\IEEEauthorrefmark{4}Faculty of Computer Science, Universitas Indonesia, Kampus Baru UI Depok, Jawa Barat, Indonesia}
		\IEEEauthorblockA{\IEEEauthorrefmark{5}Dept of Mathematics and Computer Science, Faculty of Science, University of Dschang, Cameroon}
	}
	

	

	%

	
	\maketitle
	
	\begin{abstract}
		The utilization of cloud environments to deploy scientific workflow applications is an emerging trend in scientific community. In this area, the main issue is the scheduling of workflows, which is known as an NP-complete problem. Apart from respecting user-defined deadline and budget, energy consumption is a major concern for cloud providers in implementing the scheduling strategy. The types and the number of virtual machines (VMs) used are determinant to handle those issues, and their determination is highly influenced by the structure of the workflow. In this paper, we propose two workflow scheduling algorithms that take advantage of the structural properties of the workflows. The first algorithm is called Structure-based Multi-objective Workflow Scheduling with an Optimal instance type (SMWSO). It introduces a new approach to determine the optimal instance type along with the optimal number of VMs to be provisioned. We also consider the use of heterogeneous VMs in the Structure-based Multi-objective Workflow Scheduling with Heterogeneous instance types (SMWSH), to highlight the algorithm's strength within the heterogeneous environment. The simulation results show that our proposal produces better energy-efficiency in 80\% of workflow/workload scenarios, and save more than 50\% overall energy compared to a recent state-of-the-art algorithm.
	\end{abstract}
	\IEEEoverridecommandlockouts
	
	%
	\IEEEpeerreviewmaketitle

	\section{Introduction}
	\label{Introduction}
	Energy consumption has been a major issue in cloud environments. That high energy consumption has been traced from several sources, among which the servers are the main power consumers \cite{beloglazov2011taxonomy}\cite{duy2010performance}\cite{greenberg2008cost}. In early 2010, researchers reported that under-utilization of cloud resources \cite{buyya2013introduction}\cite{reiss2012heterogeneity}\cite{beloglazov2011taxonomy} contributes to the high energy consumption in cloud data centers, which is caused by the inefficient scheduling allocation of servers resources \cite{beloglazov2011taxonomy}\cite{piraghaj2017survey}. Many works have been carried out to improve the energy-efficiency of clouds, from idle servers switching off \cite{duy2010performance}, VMs/workload consolidation \cite{garg2020energy}\cite{dabbagh2015toward}\cite{beloglazov2012energy}, to the Dynamic Voltage and Frequency Scaling (DFVS) \cite{garg2019reliability}\cite{li2015cost}\cite{tang2014efficient}\cite{guerout2013energy} techniques. 
	
	As commercial clouds gradually emerge as promising environments for the execution of scientific workflow applications on many domains such as biology, physics, medicine, and astronomy \cite{deelman2008cost}\cite{jackson2010performance}\cite{madduri2015globus}, it creates more complex challenges in term of workflow scheduling. A workflow may contain hundreds or thousands of interdependent tasks which are executed under a permanent dependency constraint where a task can only start its execution if the executions of its parents are completed. That means, allocated cloud resources can hardly be fully utilized as they might be inevitable gaps between tasks execution. This challenge brings together the complexity of clouds management along with the workflow structures complexity. The level of difficulty of cloud workflow scheduling becomes very high, especially if we consider multiple user defined Quality of Service (QoS) parameters (e.g., deadline, cost). It is very important to correctly determine the types and the number of cloud resources \cite{ndamlabin2021dynamic}\cite{rodriguez2018scheduling} to avoid the energy wastage and the violation of Service Level Agreement (SLA). No matter the types of resources being used, it is highly inefficient for the cloud providers to provision VMs in a larger amount than the largest number of tasks of the workflow's width for its execution \cite{ndamlabin2021dynamic}. Therefore, It is crucial to determine the optimal number and the types of VMs to use during the execution of a workflow, to avoid resource wastage and to ensure the SLA fulfilment as well as providing a reasonable profit for the cloud providers. A good prediction of number and type of resources for the execution of a workflow is a good asset for the designing the energy-efficient techniques.
	
	To the best of our knowledge, only few works \cite{ndamlabin2021dynamic} focused on determining suitable instance types set or a number of VMs instances in advance with an analytical approach. Some solutions use a \textit{naive determination approach} in which it is at the end that one realize which types and the number of VMs have been used, leading in more cases to a wastage (too many provisioned VMs that are less utilized). Others are time consuming determination approaches, like greedy determination \cite{faragardi2019grp} and paths-based clustering determination \cite{garg2019reliability}\cite{singh2018novel}. The paths-based clustering approach is better than the greedy one, however, its complexity and effectiveness are compromised if the workflow graph is strongly connected. Moreover, most of the solutions in the literature are effective only for a few types of workflow, while the types and structures of workflow are very complex and varied \cite{juve2013characterizing}. This is not conform with the recommendation \cite{juve2013characterizing}\cite{rodriguez2018scheduling} of designing scheduling strategies that are effective no matter the type of workflow.
	
	We advocate that homogeneity can produce better results if the good instance is chosen for the workflow execution. We further advocate that the provisioning of a correct number of suitable VMs can help not only to produce better results, but also to improve the VM utilization and further reduce the energy consumption.
	
	In this paper, we introduced a new technique based on structural properties of workflows and user-defined budget and deadline, to determine an "optimal instance type" along with the "optimal number" of VMs for adequate provisioning. In our strategy, all of the VMs used are of the same type (i.e., optimal instance type), and their number is limited by the "optimal number". Two scheduling algorithms are proposed, Structure-based Multi-objective Workflow Scheduling with an Optimal instance type (SMWSO), and Structure-based Multi-objective Workflow Scheduling with Heterogeneous instance types (SMWSH). Both algorithms aim at minimizing energy consumption, makespan, and execution cost under user-defined deadline and budget. The effectiveness of our algorithms have been proved through comparative simulations against Reliability and Energy Efficient Workflow scheduling (REEWS), a recent state-of-the-art algorithm \cite{garg2019reliability}. The choice of REEWS is because it used a clustering technique to determine the number of VMs to be used and the DVFS to minimize energy consumption.
	
	The remaining sections of the paper are organized as follows. Section 2 presents related work. In Section 3 we define the workflow scheduling problem. Section 4 describes the proposed scheduling heuristics, while Section 5 presents the experimental setup and analyses the results of the simulation experiments. Section 6 summarizes and concludes the paper.
	
	\section{Related Work}
	\label{sec:Related_Work}
	
	One of the most widely used techniques in heuristic workflow scheduling is the list-based scheduling, represented by the Heterogeneous Earliest Finish Time (HEFT) proposed by Topcuoglu et \textit{al.} \cite{topcuoglu2002performance}. HEFT aims to minimize the makespan of workflow execution in heterogeneous environments. It sorts the tasks of the workflow into a scheduling list and then assigns each task to the resource which can finish it the earliest. The determination of the finish time takes into account the already mapped tasks to each VM and find (with insertion base) the first time slot that can accommodate the task according to the precedence constraints in the workflow.
	
	In \cite{ndamlabin2020cost}, the Cost-Time Trade-off Workflow Scheduling (CTTWS) algorithm has been proposed, which aims at minimizing execution cost and execution time under user-defined budget and deadline. The CTTWS algorithm uses a trade-off function between cost and time, combined with the Implicit Requested Instance Types Range (IRITR) which is a technique aiming at determining a suitable (sub-)set VMs instance types to be used in order to avoid overbidding and underbidding. CTTWS has proved to be better to meeting users' deadlines and budgets up-to 38.4\% according to the variety of available instance types compared to the state-of-the-art. However, CTTWS rely on static VM provisioning, and failed to produce good results for SIPHT workflow due to its structure. A dynamic version of CTTWS is proposed in \cite{ndamlabin2021dynamic}, denoted Cost-Time trade-off workflow scheduling with dynamic provisioning (CTTWSDP). The CTTWSDP algorithm introduces a structure inspired dynamic VM selection and limitation. It improves the IRITR and limits the number of VMs to provision to the average width of the workflow. Through comparative simulation using five wide spread workflows, each with five different size of tasks, CTTWSDP has proved to be more effective than four state-of-the-art algorithms. Moreover, CTTWSDP is significantly effective no matter the type and the workload of workflow. 
	
	Among the algorithms proposing energy-efficient techniques we include Kimura et \textit{al.} \cite{kimura2006emprical} that proposed a slacking algorithm that uses the non-critical path to extends the task execution time by reclaiming slack time to save energy. Huang et \textit{al.} \cite{huang2012enhanced} present an enhanced Energy-Efficient Scheduling (EES) algorithm which reduces energy consumption while meeting the performance-based requirements. Then, Durillo et al. \cite{durillo2012moheft} proposed an extended version of the HEFT algorithm denoted multi-objective heterogeneous earliest finish time (MOHEFT) aiming at providing suitable trade-offs between makespan and energy consumption. 
	
	Furthermore, Huang et \textit{al.} \cite{huang2012enhanced} also proposed two algorithms extending the HEFT algorithm by introducing the energy awareness, called the Enhancing Heterogeneous Earliest Finish Time (EHEFT) and the Enhancing Critical Path on a Processor (ECPOP), and addressed the time and energy-efficient workflow scheduling. Tang et \textit{al.} \cite{tang2014efficient} introduce the DVFS enabled Efficient energy Workflow Task Scheduling (DEWTS) algorithm to obtain more energy reduction. However, the last two algorithms reserve a set of VM instances for the whole makespan.
	
	Then, Li et al. \cite{li2015cost} proposed cost and energy-aware scheduling (CEAS) algorithm to minimize the execution cost of workflow and reduce the energy consumption while meeting the deadline constraint in the cloud environment. CEAS first uses a VM selection algorithm that applies the concept of cost-utility to map tasks to their optimal virtual machine (VM) types by the sub-makespan constraint. Afterwards, it employs two tasks merging methods to reduce execution cost and energy consumption. In order to reuse the idle VM instances which have been leased, it further proposed a VM reuse policy. And finally, it utilized a scheme of slack time reclamation to save energy of leased VM instances.
	
	More recently, Ritu Garg et al. \cite{garg2019reliability} proposed the Reliability and Energy Efficient Workflow scheduling (REEWS) algorithm. The aim of their proposal is to minimize the energy consumption and maximize the reliability of the workflow execution in the respect of the user-specified QoS/deadline constraint. The REEWS  algorithm consists of four main steps: the prioritization of the tasks; the tasks clustering; (user-defined) deadline distribution among the workflow tasks and the mapping of cluster tasks to processors at suitable voltage/frequency levels in order to maximize the overall reliability of the system and minimize of energy consumption.
	
	Singh et al. \cite{singh2019energy} proposed a meta-heuristic called energy efficient workflow scheduling (EEWS) algorithm, aiming at minimize makespan and energy consumption. EEWS is inspired from hybrid chemical reaction optimization (HCRO) algorithm, and adds a new operator called on-wall pseudo-effective collision to exploit the benefits of swap mutation, and consider dynamic voltage scaling (DVS) along with a novel proposed measure to calculate the amount of energy that can be conserved.
	
	Neha Garg et al. \cite{garg2020energy} proposed the energy and resource efficient workflow scheduling (ERES) algorithm, which aims at minimizing energy consumption, maximizing resource utilization, and minimizing workflow makespan. The ERES algorithm uses VM migration to deploy/un-deploy the VMs based on the workflow task’s requirements and a double threshold policy to perceive the server’ status (overloaded/underloaded or normal). ERES also makes use of the DVFS technique.
	
	\section{Modelling of the Multi-objective Workflow Scheduling Problem}
	\label{Modelling}
	
	In this section, we present the cloud resource model, the power and energy models, the workflow model, and the problem formulation. The meanings of the parameters found throughout this paper are summarized in Table \ref{tab_notations}.
	
	\begin{table}[!t]
		\vspace{-1mm}
		\caption{Parameters Notation} \label{tab_notations}
		\vspace{-5mm}
		\begin{center}
			\resizebox{\linewidth}{!}{\begin{tabular}{ll}
					\hline
					\textbf{Notation} & \textbf{Description} \\
					\hline
					$G$ & workflow DAG \\
					$WT$ & Set of tasks of $G$\\
					$t_i$ & $i^{th}$ task of $WT$\\
					$levelWidth(l)$  & number of tasks in the level $l$\\
					$\delta$ & The user-defined deadline \\
					$\delta_i$ & The sub-deadline of the task $t_i$ \\
					$B$ & The user-defined budget \\
					$VMIT$ & Set of VMs Instances Types \\
					$vmit_k$ & The instance type $k$ of VM \\
					$VMS$ & Set of leased VMs \\
					$vm_p$ & The $p^{th}$ leased VM \\
					$p_k$ & The computing performance of $vmit_k$ \\
					$\tau$ & The length of the billing period \\
					$Cost_G$ & The total cost of executing a workflow\\
					$minM_G$ & The minimum possible makespan \\
					$minCost_G$ & The minimum possible execution cost \\

					$c_k$ & cost per billing period for $vmit_k$\\
					$Z_i$ & the weight of $t_i$ in millions of instructions\\
					$s_{ij}$ & Size of Data Transferred from $t_i$ to $t_j$ (in MB)\\
					$ET(i,k)$ & Execution time of $t_i$ on a $vmit_k$ VM\\
					$TT(i,j)$ & Data Transfer time from $t_i$ to $t_j$\\
					$EC(i,k)$ & Execution cost of $t_i$ on a $vmit_k$ VM\\
					$EST(t_i)$ & Earliest Start Time of $t_i$\\
					$EFT(t_i)$ & Earliest Finish Time of $t_i$\\
					$AST(t_i)$ & Actual Start Time of $t_i$ \\
					$AFT(t_i)$ & Actual Finish Time of $t_i$ \\
					$OCCW(t_i)$ & Sum of $TT(i,j)$, $t_j  \in Succ(t_i)$ \\
					$outd(t_i)$ & Number of immediate child of $t_i$ \\
					$\beta$ & The communication bandwidth \\
					$M_G$ & The workflow completion time (makespan)\\
					$map(i)$ & The VM on which $t_i$ is mapped to\\
					\hline
			\end{tabular}}
		\end{center}
		\vspace{-4mm}
	\end{table}
	
	\subsection{Cloud computing model}
	\label{subsec:ProposedMOAlgo_Cloud_model}
	
	The cloud data center model is similar to the one offered by Amazon EC2 \cite{AmazonEC2ResModel}. We assume that the data center is equipped with a set of $K$ types of heterogeneous VM instances, denoted by VMIT$ = \{vmit_1, vmit_2, ..., vmit_K\}$, having various processing costs, performances and configurations. Each instance type $vmit_k$ is defined by its computing performance $p_k$ in millions instructions per second (MIPS), its processing cost per billing period $c_k$ and communication bandwidth $b_k$. An instance with a higher computing performance is . For sake of simplicity, we assume that the communication bandwidth between the instances is uniformly distributed, and denoted by $\beta$.
	
	We consider that $P$ VMs ($VMS = \{vm_1, vm_2, ..., vm_P\}$), each been of an instance type in VMIT, are leased as subscription-based services in a pay-per-use model and are charged per billing period of length $\tau$. A billing period is one hour per VM usage for most IaaS providers; each partial hour consumed being rounded up to a full hour, such that 1 hour and 1 minute (61 min.) will be considered as 2 hours (120 min.) of utilization. As it is often the case \cite{AmazonEC2Pricing}. We assume that all the instance types are ordered according to their characteristics (see Eq. (\ref{eq_perf_ordering}) and (\ref{eq_cost_ordering})).
	\begin{equation}
		\label{eq_perf_ordering}
		p_1 < p_2 < ... < p_k < ... < p_K,
	\end{equation}
	
	\vspace{-0.75cm}
	
	\begin{equation}
		\label{eq_cost_ordering}
		c_1 < c_2 < ... < c_k < ... < c_K,
	\end{equation}
	
	\subsection{Power and Energy models}
	\label{subsec:ProposedMOAlgo_Energy_model}
	
	In terms of energy consumption among system components, processors consume typically the largest portion \cite{beloglazov2012energy}. Hence, we will focus on energy consumption of processors. 
	A processor consumes energy either idle or while running a task. The power consumed by a processor $p_k$ during its runtime, noted $P^k$, is expressed by equation (\ref{eq_power_consumption_k}) \cite{beloglazov2012energy}\cite{stavrinides2018impact}.
	\begin{equation}
		\label{eq_power_consumption_k}
		P^k (u_k(t)) = P^k_{idle} + (P^k_{max} - P^k_{idle}) \times u_k(t),
	\end{equation}
	
	\noindent where $P^k_{idle}$ and $P^k_{max}$ are the power consumed by the processor when idle and at 100\% utilization respectively, whereas $u_k(t)$ is the utilization rate of the processor, which is a function of the time. Therefore, the total energy consumption of a processor $p_k$ over a period of time [t0, t1] can be defined as an integral of the power consumption function over the same period (see Eq. (\ref{eq_energy_consumption_k})).
	\begin{equation}
		\label{eq_energy_consumption_k}
		E^{k} = \int_{t1}^{t0} P^k (u_k(t))  dt
	\end{equation}
	
	\noindent where the overall energy consumption ($E_{total}$) on all the $P$ VMs is simply the sum of all the energy consumption. We also assume that the DVFS is enabled on the system.
	
	\subsection{Workflow model}
	\label{subsec:ProposedMOAlgo_Workflow_model}
	
	The most commonly used model for scientific application is workflow represented as a Directed Acyclic Graph (DAG). A DAG is a graph $G (WT, E)$ where $WT = \{t_1, t_2, ... , t_n\}$ is the set of the tasks of the workflow (the weight of task $t_i$, in terms of millions of instructions, is denoted by $Z_i$), and $E = \{e_{i,j} = (t_i, t_j) | 1 \leq i,j \leq n, i \neq j \}$ a the set of edges representing the existing data and control dependencies between tasks. Thus $e_{i,j} \in E$ if there is a precedence constraint between $t_i$ and $t_j \in WT$, such that the execution of $t_j$ can start only after $t_i$ finishes its execution and sends data (of size $s_{ij}$ in Megabytes (MB)) to $t_j$. Task $t_i$ is a parent of $t_j$ and $t_j$ a child of $t_i$. A task is ready to start its execution when all of its parents have been executed and all its required data have been provided. Any task with no parent is an entry task and any task with no child is an exit task. The set of the parents (resp. children) of the task $t_j$ is denoted $pred(t_i)$ (resp. $succ(t_i)$). The Critical Path (\textbf{CP}) of a DAG is defined as its longest path.
	
	A workflow can have one or more entry tasks (tasks without parent), and one or more exit tasks (tasks without child). Entry tasks and exit tasks are denoted as $t_{entry}$ and  $t_{exit}$ respectively.
	
	Given a task $t_i$, its execution time on a resource of instance type $vmit_k$ is denoted by $ET(i,k)$ and defined as
	\begin{equation}
		\label{eq_task_execution_time}
		ET(i,k) = \dfrac{Z_i}{p_{k}},
	\end{equation}	
	
	\noindent and the data transfer time from $t_i$ to $t_j \in succ(t_i)$ denoted by $TT(i,j)$ is defined as
	\begin{equation}
		\label{eq_task_data_transfer_time}
		TT(i,j) = 
		\left\lbrace 
		\begin{array}{ll}
			0&, \ if \ t_i \ and \ t_j \ are \ mapped \ of \ the \ same \ VM \\
			\dfrac{s_{ij}}{\beta}&, \ otherwise
		\end{array}\right.
	\end{equation}
	
	The execution cost of task $t_i$ on a resource of instance type $vmit_k$ is denoted by $EC(i,k)$ and defined as
	\begin{equation}
		\label{eq_task_execution_cost}
		EC(i,k) = \left\lceil \dfrac{ET(i,k)}{\tau} \right\rceil \times c_k, 
	\end{equation}
	
	\noindent where $\tau$ is the length of a billing period. 
	
\begin{figure}
		\begin{center}
			\includegraphics[width=1.1\linewidth]{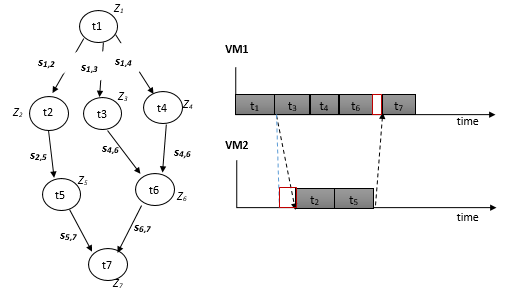}
			\caption{Example of mapping of DAG tasks onto VMs, with idle time slots due to data transfer}
			\label{fig_dag_sample_with_mapping}
			\vspace{-4mm}
		\end{center}
		\vspace{-0.4cm}\end{figure}

	Assuming that a number of tasks ($t_i$, $i \in I$) are mapped onto the same VM instance $vm_p$, supposed to be of instance type $vmit_k$, the cumulative execution cost will be evaluated as follow:
	\begin{equation}
		\label{eq_set_of_tasks_execution_cost}
		EC(I,k) = \left\lceil \dfrac{\sum_{i \in I}^{} \{ ET(i,k) \} + TITS_p }{\tau} \right\rceil \times c_k,
	\end{equation}
	
	\noindent where $TITS_p$ (Transfer Idle Time Slots) is the sum of the idle time slots due to data transfer awaited by the VM $vm_p$. The example presented in Figure \ref{fig_dag_sample_with_mapping} gives an illustration of TITS for the mapping of a DAG of seven tasks onto two VMs. The red rectangles are TITS.
	
	We denote the Earliest Start Time and Earliest Finish Time of task $t_i$ as $EST(t_i)$ and $EFT(t_i)$ respectively. They are defined as
	\begin{equation}
		\label{eq_task_earliest_start_time}
		EST(t_i) = 
		\left\lbrace 
		\begin{array}{ll}
			0&, \ if \ t_i = t_{entry}, \\
			\max_{}^{}{ \{ EFT(t_j) +TT(j,i) \} }&, \ otherwise\\
			^{t_j \in pred(t_i)}&
		\end{array}\right.
	\end{equation}
	
	\noindent and
	\begin{equation}
		\label{eq_task_earliest_finish_time}
		EFT(t_i) = EST(t_i) + ET(i,K), 
	\end{equation}
	
	\noindent where $ET(i,K)$ is the minimum possible $ET$ of $t_i$ over all possible VM instance types.
	
	In like manner, we also consider the Actual Start Time and the Actual Finish Time of a task $t_i$ mapped onto a VM $vm^k_p$ of instance type $vmit_k$. They are denoted by $AST(t_i,vm^k_p)$ and $AFT(t_i,vm^k_p)$ respectively. Their values may be different from those of $EST(t_i)$ and $EFT(t_i)$, due to the heterogeneity of VMs and the fact that functions EST and EFT provide the minimum possible value over all possible mapping of tasks onto VMs. The workflow completion time (also called makespan), denoted by $M_G$, is defined as the Actual Finish Time of the exit task $t_{exit}$. The minimum makespan, denoted by $minM_G$, is defined as the Earliest Finish Time of $t_{exit}$. Thus, $minM_G = EFT(t_{exit})$.
	
	Finally, the total cost of executing a workflow is the sum of the execution cost of all the tasks when effectively mapped onto VMs, and defined as: 
	\begin{equation}
		\label{eq_workflow_total_execution_cost}
		Cost_G = \sum_{t_i \in G}^{} EC(i,map(i)),
	\end{equation}
	
	\noindent where $map(i)$ denotes the VM on which task $t_i$ is mapped, $1 \leq p \leq P, \ 1 \leq k \leq K$.
	
	\subsection{Problem Formulation}
	\label{subsec:ProposedMOAlgo_Problem_formulation}
	
	The role of a workflow scheduler is to determine an execution order of the workflow tasks, and the VM onto which to assign each task. That mapping of tasks onto VMs have to satisfy some requirements of the user and the cloud provider.
	
	In this paper, the targeted objectives of the workflow scheduler are the minimization of the overall execution cost and execution time, as well as the energy consumption of the system. The constraints are the user-defined deadline ($\delta$) and the user-defined budget ($B$).
	
	
	The problem can then be formulated as follow: \textit{how to build a workflow scheduling algorithm, able to dynamically provision VMs for tasks execution in order to minimize the overall execution cost and execution time as well as the energy consumption of the system, under of the user-defined budget and deadline?}
	
	The problem can be formulated as a mathematical optimization problem:
	
	\begin{equation}
		\label{eq_pb_formulation}
		\left\lbrace 
		\begin{array}{c}
			Minimize (E_{total})\\
			Subject \ to \ M_G \leq \delta \ and \ Cost_G \leq B
		\end{array}\right.
	\end{equation}

	%
	%
	\section{The Proposed Scheduling Algorithms}
	\label{SMWS_algo_desc}
	
	\begin{figure}[!t]
		\begin{center}
			\includegraphics[width=\linewidth]{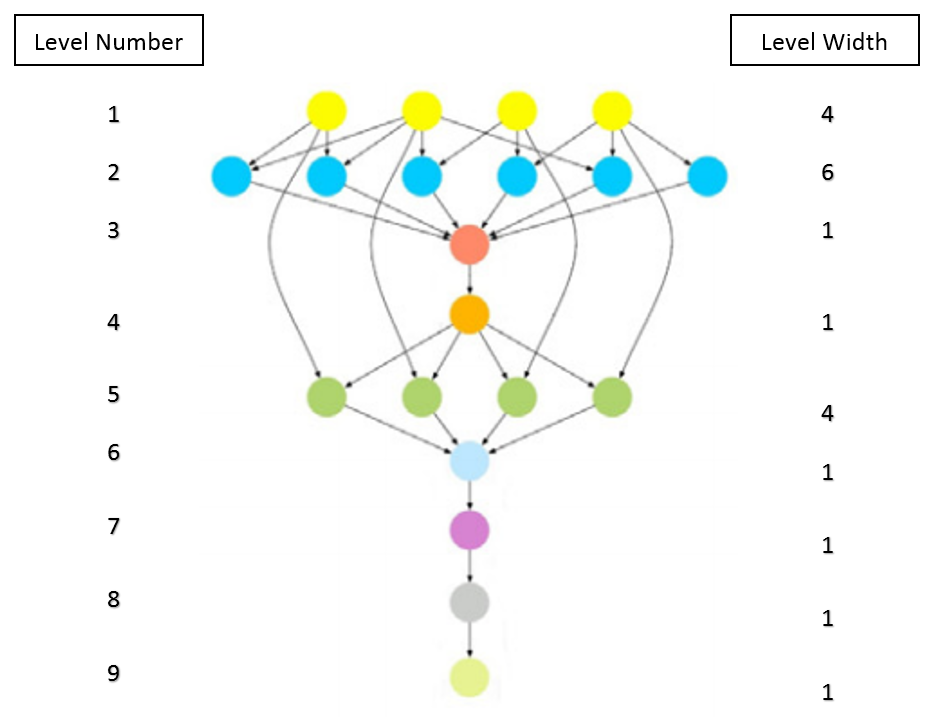}
			\caption{Workflow width distribution.}
			\label{fig_workflow_widths_smws}
		\end{center}
		\vspace{-0.4cm}\end{figure}
	
	In this section, we present our proposed solutions for the workflow scheduling problem. Here we are presenting two algorithms, Structure-based Multi-objective Workflow Scheduling with an Optimal instance type (SMWSO), and Structure-based Multi-objective Workflow Scheduling with Heterogeneous instance types (SMWSH). Both algorithms are dynamic and rely on dynamic provisioning. First of all, we present some scheduling techniques used in our proposals, and then their whole description.
	
	\subsection{Determination of the Optimal number of VMs and the Optimal Instance type}
	\label{SMWS_algo_desc:optimalNumberAndInstance}
	
	Figure \ref{fig_workflow_widths_smws} shows the structure of a Montage workflow with twenty tasks and their dependencies. In this figure, the left column shows level numbers calculated by equation (\ref{eq_wf_level_number_1}), while the right column is the number of tasks in each level that we call the level width ($levelWidth(l)$). In this example the largest width is 6 which corresponds to the level 2 ($levelWidth(2) = 6$).
	\begin{equation}
		\label{eq_wf_level_number_1}
		LN (t_j) = 1 + \max_{t_i \in pred(t_j)}^{}{ \{ LN (t_i) \} },
	\end{equation}
	
	Obviously, it is unlikely to use more VMs than the largest width of the workflow for its execution. But at the same time how many VMs is suitable for the execution of the workflow? In \cite{ndamlabin2021dynamic}, we have proposed the average width as suitable number of VM to provisioned for the execution of the workflow. In this paper we are using the standard deviation and the maximum width along with the average width (see Eq. (\ref{eq_optimal_number_vms})).
	
	{\footnotesize
		\begin{equation}
			\label{eq_optimal_number_vms}
			ONVM(wf) = 
			\left\lbrace 
			\begin{array}{ll}
				AvgW_{Wf}, \ if \ AvgW_{Wf} \leq StdDW_{Wf} & \\
				\min_{}^{}{ \{AvgW_{Wf} + StdDW_{Wf} , MaxW_{Wf}\} }, \ otherwise &\\
			\end{array}\right.
		\end{equation}
	}
	\noindent where $MaxW_{Wf}$ is the maximum, $AvgW_{Wf}$ the average and $StdDW_{Wf}$ the standard deviation of the levels' width of the workflow. For the example of Figure \ref{fig_workflow_widths_smws}, the determination of the optimal number of VMs is presented in Table \ref{tab_optNbVMs_determination_exemple}.
	
	\begin{table}[!t]
		\centering
		\caption{Example of determination of optimal number of VMs based on Figure \ref{fig_workflow_widths_smws}.} \label{tab_optNbVMs_determination_exemple}
		
		\resizebox{\linewidth}{!}{\begin{tabular}{c c c c c }
				\hline
				\textbf{Width distribution}  & \textbf{Max} & \textbf{Avg} & \textbf{Std Dev.} & \textbf{optimal Nb of VMs} \\ 
				\hline
				\textbf{4; 6; 1; 1; 4; 1; 1; 1; 1} & $6$ & $2.22$ &  $1.81$  & $2.22 + 1.81 \simeq 4$  \\
				\hline
		\end{tabular}}
		\vspace{-4mm}
	\end{table}
	
	It is established that instances heterogeneity can easily leads to more energy wastage, due to the workload unbalance of instances. For example, Stavrinides and Karatza \cite{stavrinides2018impact} studied the impact of the workload and their results reveal that the workload variability has a significant impact on the energy consumption of the system. We also investigated and found that if no careful VM selection is made, using different instances for the execution of a workflow leads to more energy wastage than when one suitable instance is chosen.
	
\begin{figure}[!t]
		\begin{center}
			\includegraphics[width=\linewidth]{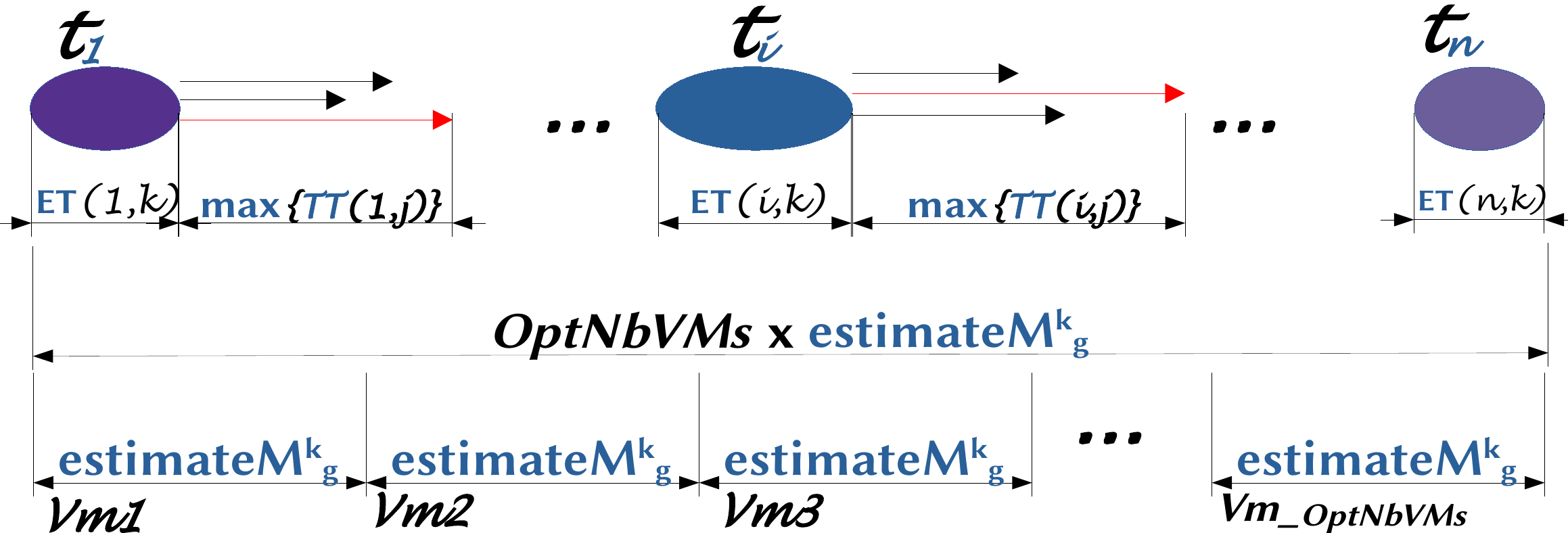}
			\caption{Makespan estimation when using an unique instance type $vmit_k$}
			\label{fig_Estimated_Makespan_vmitK}
		\end{center}
		\vspace{-4mm}
		\vspace{-0.4cm}\end{figure}

	Since the optimal number of VMs ($optNbVMs$) is determined regardless the type of instance, we can then chose the instance type ($optInstType$) which gives better results. The effectiveness of this operation is highly dependent on the estimation of the makespan. Our makespan estimation when using only one instance type $vmit_k$ ($estimateM_G^k$) is made according to equation (\ref{eq_makespan_estimation_single_instance_type}). 
	\begin{equation}
		\label{eq_makespan_estimation_single_instance_type}
		estimateM_G^k = \dfrac{\sum_{t_i \in WT}^{} \{ET(i,k) + \max_{t_j \in Succ(t_i)} \{ TT(i,j) \}  \}   }{optNbVMs}
	\end{equation}
	
	\noindent where $TT(i,j)$ is the transfer time of data from task $t_i$ to task $t_j$ An illustration of that estimation is given by Figure \ref{fig_Estimated_Makespan_vmitK}.
	
	Since their is a parallelism according to the workflow structure, $optInstType$ is chosen as the fastest instance type ($vmit_k$) which has an estimated makespan that respects the deadline, with an execution cost lest than a slice of the budget corresponding to a path ($B / optNbVMs$). That means, the instance $vmit_k$ which respects conditions (\ref{eq_in_the_deadline_condt}) and (\ref{eq_in_the_budget_condt}).
	\begin{equation}
		\label{eq_in_the_deadline_condt}
		estimateM_G^k \leq \delta
	\end{equation}
	
	\vspace{-0.7cm}
	
	\begin{equation}
		\label{eq_in_the_budget_condt}
		\lceil estimateM_G^k / \tau \rceil \times c_k \leq (B / optNbVMs)
	\end{equation}
	
	That is the fastest instance type which can respect both the deadline and the budget, according to our makespan estimation and based on the optimal number of VMs ($optNbVMs$).
	
	\subsection{Deadline distribution}
	\label{SMWSH_algo_desc:Deadline_distribution}
	
	Since $minM_G$ is the minimum possible makespan of the workflow $G$, we can assume the user defined deadline is always greater than $minM_G$ ($\delta \geq minM_G$). Let be $EST^{opt}(t_i)$ (respectively $EFT^{opt}(t_i)$) the Earliest Start Time (respectively Earliest Finish Time) of task $t_i$ when executed on $optInstType$. We define the sub-deadline $\delta_i$ of each task $t_i$ as follow:
	\begin{equation}
		\label{eq_task_sub_deadline_step1}
		\delta_i^{'} = \dfrac{EFT^{opt}(t_i) \times \delta}{minM_G} ,
	\end{equation}
	
	\vspace{-0.7cm}
	
	\begin{equation}
		\label{eq_task_sub_deadline}
		\delta_i = \delta_i^{'} + spare\delta_i  ,
	\end{equation}
	
	\noindent where $spare\delta_i$ is obtain by distributing the eventual spare time ($\delta - \max_{t_i \in WT}^{} \{ \delta_i^{'} \}$) to all the task proportionally to their length compared to that of the CP.
	
	\subsection{Task priority}
	\label{SMWS_algo_desc:tasks_ordering}
	
	The order of execution of workflow tasks is very important in a scheduling strategy. Any ordering strategy used for workflow tasks most take into account the precedence constraints between the tasks. One of the most used ordering strategy is the up-rang of Topcuoglu et \textit{al.} \cite{topcuoglu2002performance}, which has been improved by Wang et \textit{al.} \cite{wang2016hsip}. In our case, the tasks prioritization strategy is an improved version of \cite{wang2016hsip} and is given by equation (\ref{eq_workflow_task_rank_u}). 
	{\small	\begin{equation}
			\label{eq_workflow_task_rank_u}
			rank_u(t_i) = \sigma_{i} + outd(t_i) + \overline{OCCW(t_i)} + \max_{{t_j \in Succ(t_i)}}^{}{ \{ rank_u(t_j) \}}
		\end{equation}
	}
	\noindent where $\sigma_{i}$ is the standard deviation of the computation time of the task $t_i$ on the available pool of processors. The task with the highest $rank_u$ is more prioritized.
	
	We are using just the standard deviation instead of multiplying it with the average computation time as it is done in \cite{wang2016hsip}. Furthermore, we are using the average communication cost weight ($\overline{OCCW(t_i)}$) instead of the OCCW, and in addition we are using the out-degree of the task which will grant more priority to tasks having more children.
	
	\subsection{VM Selection and reuse}
	\label{SMWSO_algo_desc:VM_Selection}
	
	A task $t_i$ can be mapped to a VM during the Entry Task Duplication Policy phase (see Section \ref{SMWS_algo_desc:Task_duplication}) or during the Pipeline Merging and Slacking phase (see Section \ref{SMWS_algo_desc:pipelineMergingAndSlacking}). Therefore, when a task is selected due to its priority, it is just ignored in this phase, then executed to the already mapped VM at due time.
	
	When a task $t_i$ is not yet mapped to a VM, the VM selection strategy is a modified version of the one used in HEFT \cite{topcuoglu2002performance}; ie the VM with the smallest actual finish time determined by finding the first idle time slot capable of holding the task. 
	For SMWSO, we determine among the already provisioned VMs, the one having the smallest earliest finish time, whereas for SMWSH, it is the VM having the smallest actual finish time under the sub-deadline of the task ($\delta_i$). If the corresponding start time is late on the earliest start time of the task, we proceed to the provisioning of a new VM (of type $optInstType$), taking into account the supply time. If the number of already provision VMs is up-to $optNbVMs$, we use the VM having the smallest AFT among the available VMs.
	
	if such VM doesn't exists or the corresponding start time is late on the earliest start time of the task, we proceed to the provisioning of a new VM. The new VM is of type $optInstType$ for the case of SMWSO, and for the case of SMWSH it is the instance that can end faster and under the sub-deadline ($\delta_i$). If the number of already provision VMs is up-to $optNbVMs$, we use the VM having the smallest AFT among the available VMs. The maximum number of VMs to provisioned is set to $optNbVMs$.
	
	\begin{figure}[!t]
		\begin{center}
			\includegraphics[width=\linewidth]{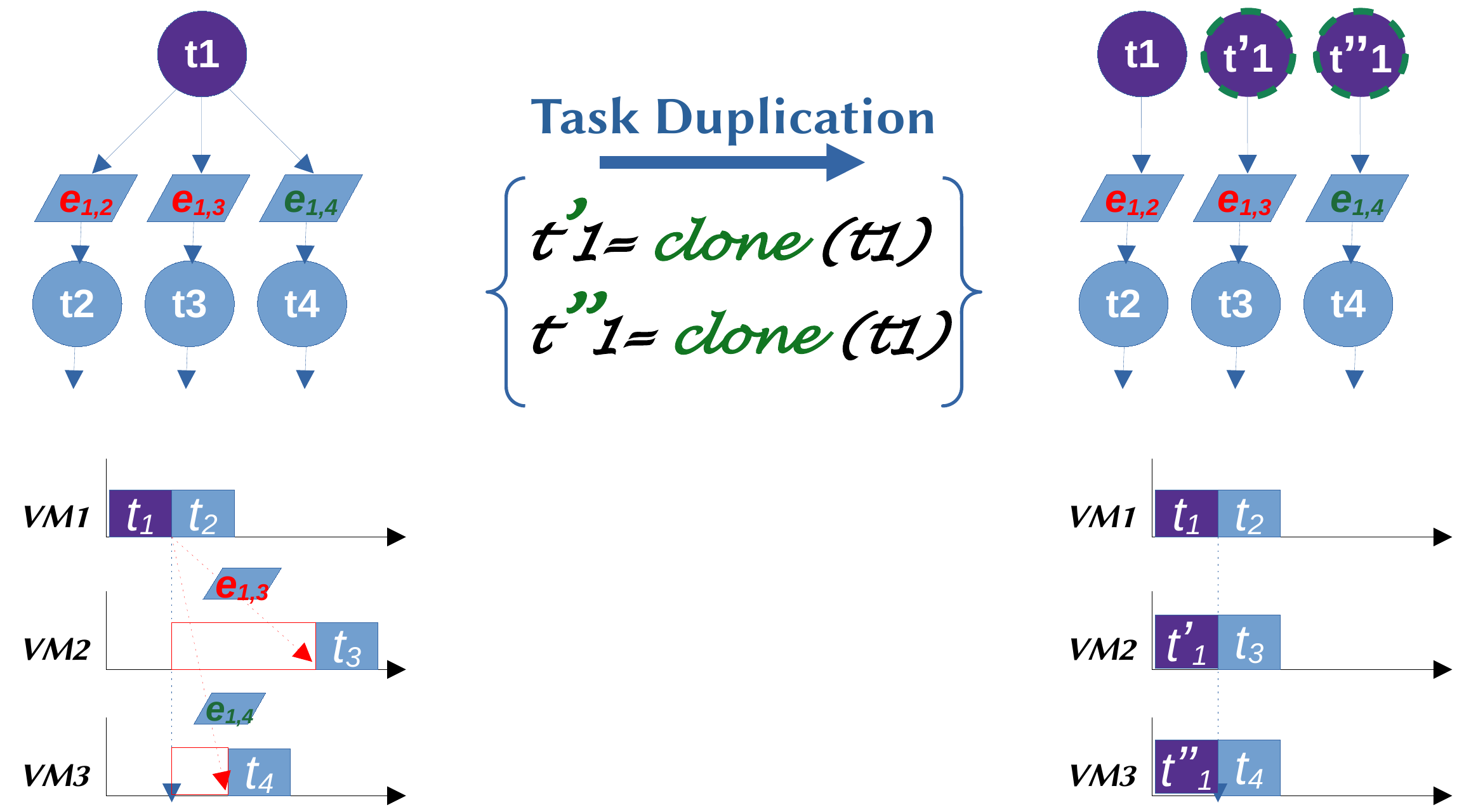}
			\caption{Entry Task Duplication}
			\label{fig_Task_Duplication}
		\end{center}
		\vspace{-0.4cm}\end{figure}
	
	\subsection{Entry Task Duplication Policy}
	\label{SMWS_algo_desc:Task_duplication}
	
	The entry task duplication is a technique used to reduce the execution time of the workflow by eliminating transfer time from the entry task \cite{wang2016hsip}. Task duplication can have an impact over the makespan, the cost and the energy consumption. To the best of our knowledge, the entry task duplication selection policies found in the literature does not take into account the case where there are several root tasks in the structure of the workflow. While it help reducing the makespan, it may raise a significant increase of the execution cost and even the energy consumption if the entry task is CPU intensive. It is then necessary to limit the number of duplication and use an accurate duplication policy.
	In our case, more than one root tasks can be duplicated according to the following policy (see Fig. \ref{fig_Task_Duplication}): 
	
	\begin{enumerate}
		\item If \textbf{$totalRepDue = min(optNbVMs,levelWidth(2))\\ - levelWidth(1) > 0$}, then there may be task duplication;	
		
		\item Under the condition (1.), an entry task $t_i$ can be duplicated if it is the only parent of more than one child\\ ($nbSingleParentChildren_i > 1$). That is, no duplication for children having several parents;

		\item Assuming the $nbSingleParentChildren_i$ children ordered according to their priority, proceed to a task duplication of the $nbSingleParentChildren_i - 1$ first children ($t_j$) as long it is possible:
		\begin{enumerate}
			\item Duplicate $t_i$ to the VM $vm_p$ ($vm_p \in VMS, vm_p <> map(i)$) with the lowest execution cost and which fulfills the condition of equation (\ref{eq_duplication_condition}) 
			\begin{equation}
				\label{eq_duplication_condition}
				ET(i,vm_p) < ET(i,map(i)) + TT(i,j)
			\end{equation}
			
			\item If a such VM $vm_p$ exists, map $t_j$ to $vm_p$ ($map(j) = vm_p$)
			\item \label{Task_duplication:If_noVM} If not, provision a new VM $vm_{p^{'}}$ of type $optInstType$, map $t_j$ to $vm_{p^{'}}$ ($map(j) = vm_{p^{'}}$)
		\end{enumerate}
		\item Assign the first unmapped child (among\\ $nbSingleParentChildren_i$ children) to $map(i)$;
	\end{enumerate}
	
	\noindent We will investigate the duplication on behalf of children having several parents in our future work.

	\subsection{Pipeline Merging and Slacking}
	\label{SMWS_algo_desc:pipelineMergingAndSlacking}
	
	The pipelines Merging and Slacking is a scheduling technique that aim at maximizing resources utilization, reducing energy consumption, and reducing execution cost eventually, through a smart management of sequential and parallel tasks.
	
	A pipeline generally starts from distribution task, ends with an aggregation task, and have a succession of tasks having exactly one parent and one child in the middle. Here, we deal with parallels pipelines, that is a set of pipelines or process tasks coming from the same parent (which is a distribution task) and leading to the same child (which is an aggregation task). 
	
\begin{figure}[!t]
		\begin{center}
			\includegraphics[width=\linewidth]{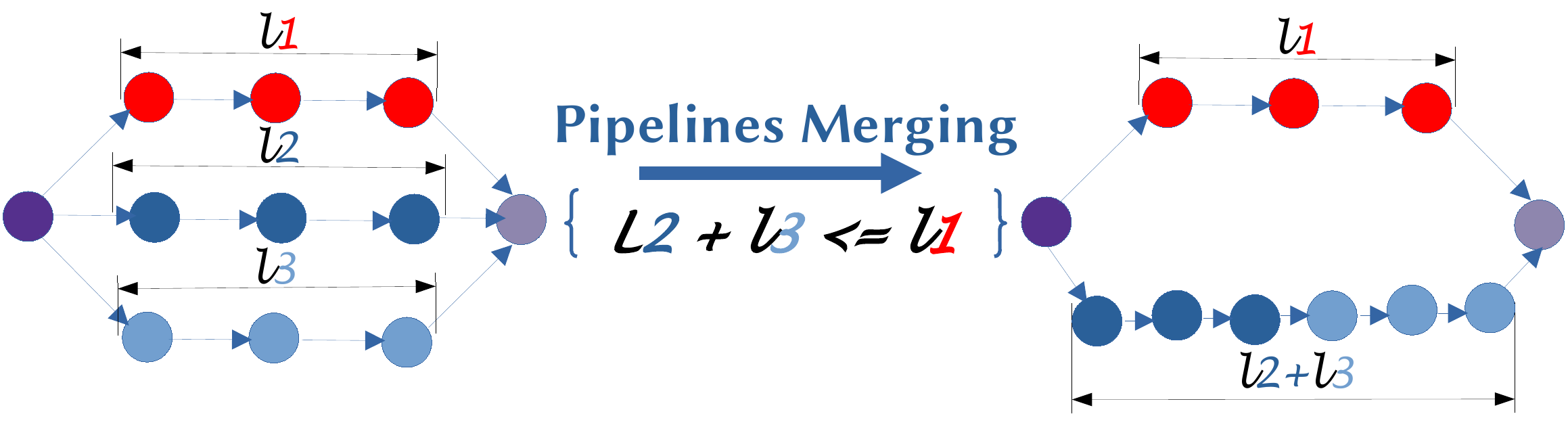}
			\caption{Pipelines Merging}
			\label{fig_Pipelines_Merging}
		\end{center}
		\vspace{-0.4cm}\end{figure}

	Our Pipeline Merging and Slacking process is in two phases (see Fig. \ref{fig_Pipelines_Merging}). The first phase is to identify parallel pipelines and merge some pipelines in the group, and the second is to apply slack time reclamation to the tasks of some pipelines in each group.
	
	\paragraph{Parallel pipelines grouping and merging:} If the current task $t_i$ is a distribution task (ei. $outd(t_i) > 1$): 
	\begin{enumerate}
		\item Determine whether there are pipelines beginning from one of its children;
		\item Construct groups of parallel pipelines;		
		\item In each group of parallel pipelines:	
		\begin{enumerate}
			\item Determine the longest pipeline (the one having the highest sum of computation length ($pipeGpL_{max}$));
			\item Constitute sub-groups of pipelines in which the sum of computation length is less or equal to the length of the longest pipeline ($pipeGpSumL \leq pipeGpL_{max}$);		
		\end{enumerate}
	\end{enumerate}
	
	The pipelines in the same sub-group could then be mapped to the same VM instance without delaying the execution time, rather, it is possible to have slack times to reclaim.  
	
\begin{figure}[!t]
		\begin{center}
			\includegraphics[width=\linewidth]{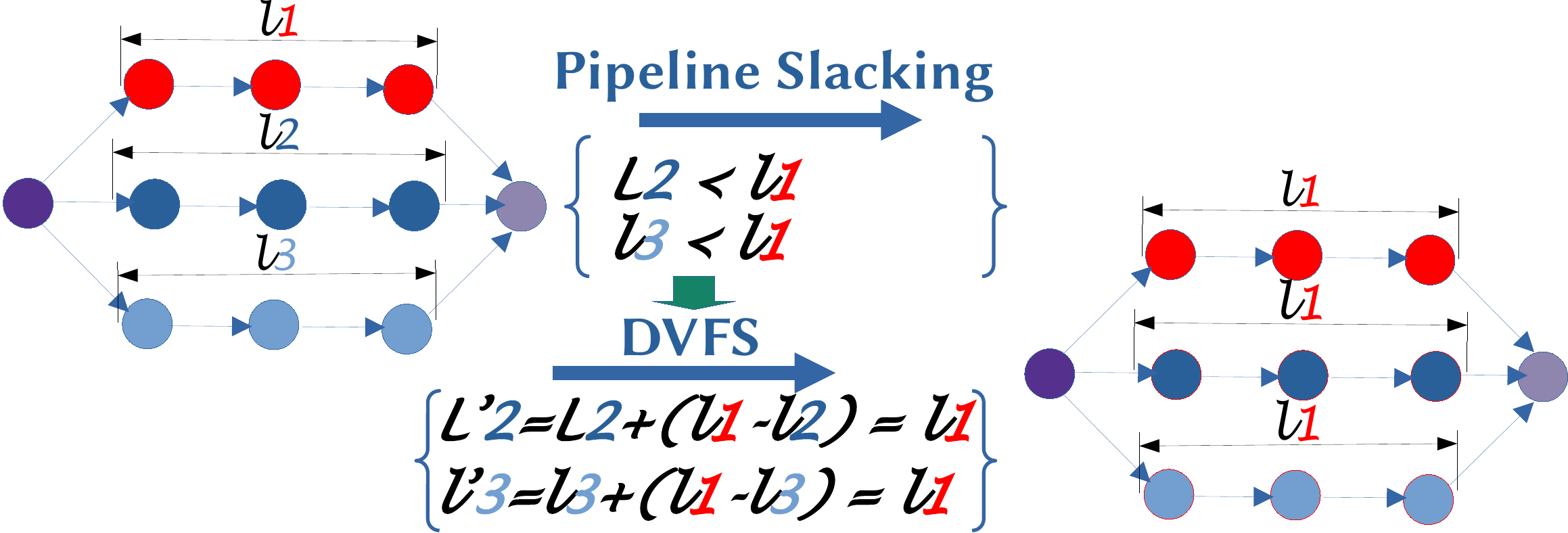}
			\caption{Pipelines Slacking}
			\label{fig_Pipelines_Slacking}
		\end{center}
		\vspace{-0.4cm}\end{figure}

	\paragraph{Pipeline Slacking:} The process of the Pipeline Slacking is as follow (see Fig. \ref{fig_Pipelines_Slacking}): if the current task $t_i$ is at the head of a pipeline  : 
	\begin{enumerate}
		\item Determine the slack time: $slackTime = pipeGpL_{max} - pipeGpSumL$;
		\item Determine the CPU utilization rate: $cpuUtilization = pipeGpL_{max} / (pipeGpL_{max} + slackTime)$;
		\item Set the CPU utilization rate of $t_i$ to $cpuUtilization$ using the DVFS technique;	
		\item For all the other tasks of the current pipeline (subsequent children of $t_i$), and all the tasks of the pipelines in the same sub-group:	
		\begin{enumerate}
			\item Map the task to $map(i)$;
			\item Using the DVFS technique, set the CPU utilization rate of the task to $cpuUtilization$;
		\end{enumerate}	
	\end{enumerate}
	
	The Pipeline Merging and Slacking technique helps in the reduction of the makespan, the energy consumption, and also the execution cost.

	\subsection{Description of the SMWSH and SMWSO algorithms}
	\label{SMWS_algo_desc:SMWS_algo_desc}
	
	In this section, we present the overall description of the two proposals, named Structure-based Multi-objective Workflow Scheduling with an Optimal instance type (SMWSO), and Structure-based Multi-objective Workflow Scheduling with Heterogeneous instance types (SMWSH). The SMWSO algorithm uses homogeneous instance according to the determination of Optimal Instance type ($optInstType$) described in section \ref{SMWS_algo_desc:optimalNumberAndInstance}, and the SMWSH algorithm uses heterogeneous instances. Both algorithms strive to enable the cloud scheduler to spend less money to complete a workflow, while minimizing makespan and energy consumption. While SMWSO consists of five main steps that are:
	
	\begin{itemize}
		
		\item The determination of the optimal instance type and the optimal number of VMs as defined in section \ref{SMWS_algo_desc:optimalNumberAndInstance};
		
		\item The task prioritization: the workflow tasks are ordered according the their descendant ranku. The ranku if defined in section \ref{SMWS_algo_desc:tasks_ordering};
		
		\item The VM selection and reuse;
		
		\item The entry task duplication as defined in section \ref{SMWS_algo_desc:Task_duplication};
		
		\item The parallels pipelines merging and slacking as defined in section \ref{SMWS_algo_desc:pipelineMergingAndSlacking}.
		
	\end{itemize}
	
	SMWSH has in addition to those five steps, the deadline distribution (see section \ref{SMWSH_algo_desc:Deadline_distribution}). The pseudo-code of SMWSO is depicted in Algorithm \ref{algo_SMWS}.

	\begin{algorithm}[!t]
		\textbf{Input:} {The DAG of tasks}\\
		\textbf{Output:} {All the tasks are scheduled to their suitable VMs}
		\caption{SMWSO Algorithm}
		\label{algo_SMWS}
		\begin{algorithmic}[1]
			\State \textbf{Starting from the $t_{exit}$, compute $rank_u$ for all tasks by using equation (\ref{eq_workflow_task_rank_u})}
			\State \textbf{Sort the tasks list ($readyList$) in decreasing order of $rank_u$}
			\State \textbf{Determine the optimal number of VMs ($optNbVMs$) and the optimal instance type ($optInstType$) as described in Section \ref{SMWS_algo_desc:optimalNumberAndInstance}}
			\For {$t_i \in readyList$}
			\State \textbf{Map $t_{i}$ to the suitable VM according to the VM selection as described in Section \ref{SMWSO_algo_desc:VM_Selection}} {\footnotesize /*If the number of the already provision VMs is up-to $optNbVMs$, we provision no more and we use the VM having the smallest AFT*/}
			\State \textbf{Apply entry task duplication over $t_{i}$ if needed as described in Section \ref{SMWS_algo_desc:Task_duplication}}
			\State \textbf{Apply pipeline merging and slacking over $t_{i}$ if needed as described in Section \ref{SMWS_algo_desc:pipelineMergingAndSlacking}}			
			\EndFor
		\end{algorithmic}
	\end{algorithm}

	The SMWSH algorithm begins with a VM of instance type $optInstType$. But unlike SMWSO, SMWSH handles VM selection, Entry Task Duplication Policy, and Pipeline Merging and Slacking as follow:
	
	\begin{enumerate}
		\item \textit{VM selection}: if the corresponding start time of the VM having the best AFT is late on the earliest start time of the task, a new provisioning is proceeded. It determines the fastest instance type that can execute the task in its sub-deadline;
		
		\item \textit{Entry Task Duplication Policy}: in the sub-step \ref{Task_duplication:If_noVM} of the task duplication policy, since the VMs are heterogeneous, it determines an instance type that can fulfils the condition of equation (\ref{eq_duplication_condition}) and provision a VM of that type.
		
		\item \textit{Pipeline Merging and Slacking}: here also because the VMs are heterogeneous, the spare time slacking takes into account the speed of the related VMs;		
		
	\end{enumerate}

	The pseudo-code of SMWSH is depicted in Algorithm \ref{algo_SMWS_Het}. 

	\begin{algorithm}[!t]
		\textbf{Input:} {The DAG of tasks}\\
		\textbf{Output:} {All the tasks are scheduled to their suitable VMs}
		\caption{SMWSH Algorithm}
		\label{algo_SMWS_Het}
		\begin{algorithmic}[1]
			\State \textbf{Starting from the $t_{exit}$, compute $rank_u$ for all tasks by using equation (\ref{eq_workflow_task_rank_u})}
			\State \textbf{Sort the tasks list ($readyList$) in decreasing order of $rank_u$}
			\State \textbf{Determine the optimal number of VMs ($optNbVMs$) and the optimal instance type ($optInstType$) as described in Section \ref{SMWS_algo_desc:optimalNumberAndInstance}}
			\State \textbf{Provision one VM of type $optInstType$}
			\For {$t_i \in readyList$}
			\State \textbf{Map $t_{i}$ to the suitable VM according to the VM selection policy} {\footnotesize /*If the number of the already provision VMs is up-to $optNbVMs$, we provision no more and we use the VM having the smallest AFT*/}
			\State \textbf{Apply entry task duplication over $t_{i}$ if needed as described in Section \ref{SMWS_algo_desc:Task_duplication}}
			\State \textbf{Apply pipeline merging and slacking over $t_{i}$ if needed as described in Section \ref{SMWS_algo_desc:pipelineMergingAndSlacking}}			
			\EndFor
		\end{algorithmic}
	\end{algorithm}

	\section{The Time Complexity of the studied algorithms}
	
	The REEWS algorithm has a time complexity order of $O(n^2)$. In fact, as a DAG has in worst case a maximum of $\frac{n(n-1)}{2}$ dependencies, and as REEWS uses a clustering technique based on paths determination, we have at most $O(n^2)$ time complexity.
	
	As for our two algorithms, the phases that must be considered are: task selection and VM selection. For task selection, we need $O(n^2)$ the Ranku, and $O (n log n)$ for tasks sorting. For the mapping, we need $O(n \times P)$. Since $P = optNbVMs < n$, SMWSO and SMWSH algorithms have a complexity of $O(n^2)$.	
	
	Therefore, the three algorithms REEWS, SMWSO and SMWSH have a polynomial time complexity of $O(n^2)$.

	\section{Performance evaluation}
	\label{sec:ProposedMOAlgo_Performance_evaluation}
	
	In this section, we present the experiment's setup and analyze the simulation results.
	
	We have used the Pegasus workflow generator \cite{bharathi2008characterization} during experimentation to create the structure of the five real-world scientific workflows (Montage, CyberShake, Epigenomics, SIPHT, and LIGO), in different workload (the number of tasks of the workflow): 50, 100, 200, 500 and 1000 tasks.

	To evaluate the performances of our two heuristics, we have implemented them as well as a state-of-the-art heuristic algorithm \cite{garg2019reliability} called Reliability and Energy Efficient Workflow scheduling (REEWS). REEWS aims at minimizing energy consumption and maximizing the reliability of the system in the respect of the user-specified deadline. The choice of REEWS is due to the fact that it uses a (clustering) technique of determination of number of VMs to use, and the DVFS. Unlike our proposals, the REEWS algorithm relies on static provisioning of VMs. We have not been able to find a single workflow scheduling heuristic, aiming at minimizing energy consumption which uses a dynamic VMs provisioning strategy. Therefore, in order to compare our two algorithms against REEWS, we have designed their static VMs provisioning versions. We have implemented our two heuristic (dynamics and statics versions) as well as REEWS algorithm \cite{garg2019reliability}. The simulations have been done in CloudSim \cite{calheiros2011cloudsim}. The performance difference between the static and the dynamic versions of each of our proposal is not significant (1\% and 7.48\%).

	\subsection{Experiment Setup}
	\label{subsec:ProposedMOAlgo_Experiment_Setup}
	
	For the simulations we consider the system as a single data center having ten different instance types that are based on the US-east (Ohio) Amazon region  \cite{AmazonEC2ResModel}, collected in July 2019, and which the characteristics are presented in Table \ref{tab_EC2_Instance_Types_With_Power}. The last two columns concerning the power were taken from the CloudSim framework \cite{calheiros2011cloudsim} modified by Guerout et \textit{al.} \cite{guerout2013energy}.
	
	\begin{table}[!t]
		\centering
		\caption{Instance types based on Amazon EC2} \label{tab_EC2_Instance_Types_With_Power}
		\footnotesize
		\resizebox{\linewidth}{!}{\begin{tabular}{ c c c c c c }
				\hline
				\multirow{2}{*}{\textbf{Type}} & \multirow{2}{*}{\textbf{vCPU}} & \multirow{2}{*}{\textbf{Memory(GB)}} & \multirow{2}{*}{\textbf{Cost(\$)/Hour}} & \multicolumn{2}{c}{\textbf{Power (W)}} \\
				\cline{5-6}
				& & & & \textbf{Min} & \textbf{Max} \\
				\hline
				m3.medium & 	1 & 3.75 & 0.067 & 140 & 228\\
				
				m4.large & 	2 & 8 & 0.10 & 146 & 238\\
				
				m4.xlarge & 	4 & 16 & 0.20 & 153 & 249\\
				
				m4.2xlarge & 	8 & 32 & 0.40 & 159 & 260\\
				
				m4.4xlarge & 	16 & 64 & 0.80 & 167 & 272\\

				m5.8xlarge & 	32 & 128 & 1.536 & 174 & 282\\
				
				m4.10xlarge & 	40 & 160 & 2.00 & 182 & 294\\
				
				m5.12xlarge & 	48 & 192 & 2.304 & 188 & 305\\
				
				m4.16xlarge & 	64 & 256 & 3.20 & 196 & 316\\
				
				m5.24xlarge & 	96 & 384 & 4.608 & 204 & 330\\		
				\hline
		\end{tabular}}
	\end{table}

	We have configured the simulation environment as follows. The bandwidth between instances is fixed to 20 MBps, the value of the vCPU of each instance is considered as its processing capacity in Million Instruction Per Second (MIPS) as in \cite{rodriguez2018scheduling}, and the charging model is hourly based. For the dynamic provisioning of VMs, the provisioning delay of each VM was set to 100 s based on the study by Mao et \textit{al.} \cite{mao2012performance}. In the case of experiment with static provisioning, we have created 10000 VMs such that the number of VMs per instance type is the same. Finally, we suppose that the DVFS is enabled on the different resource.

	We have calculated for each the fastest schedule ($FS$) as a baseline schedule and the lowest budget ($LB$) as follow:
	\begin{equation}
		\label{eq_fastest_schedule}
		FS = \sum_{t_i \in CP}^{} ET(i,K), 
	\end{equation}
	
	\noindent where $ET(i,K)$ is the execution time of task $t_i$ on the fastest instance according to equation (\ref{eq_task_execution_time}). FS can be viewed as the sum of the minimum execution times of the tasks belonging to the Critical Path (CP).
	\begin{equation}
		\label{eq_lowest_budget}
		LB = \sum_{t_i \in G}^{} EC(i,1),
	\end{equation}
	
	\noindent where $EC(i,1)$ the execution cost of task $t_i$ on the cheapest instance according to equation (\ref{eq_task_execution_cost}). LB is the lowest possible cost required for executing a workflow, irrespective of the completion time.

	Then by using equations (\ref{eq_fastest_schedule}) and (\ref{eq_lowest_budget}), we set variation ranges for user-defined budget and deadline from tight to moderate to relaxed as follow:
	\begin{equation}
		\label{eq_deadline_range}
		deadline = \alpha * FS, \alpha \in [4, 8, 12, 16],
	\end{equation}
	
	\vspace{-0.7cm}
	
	\begin{equation}
		\label{eq_budget_range}
		budget = \beta * LB, \beta \in [4, 8, 12, 16],
	\end{equation}

	For each of the five workflow and each of the five different sizes (50, 100, 200, 500, and 1000 tasks), we have conducted thirty (30) experiments ($25 \times 30 = 750$). The variation of deadline and budget factors yields 16 different cases. By considering both deadline and budget variations, the number of experiments is $750 \times 16 = 12 000$.

	\subsection{Performance Metrics}
	\label{subsec:ProposedMOAlgo_Performance_Metrics}
	
	We compare the performances of three algorithms REEWS, SMWSO and SMWSH based on the following well-known performance metrics: 
	
	\begin{figure*}[!t]
		\centering
		\begin{subfigure}[t]{.325\linewidth}
			\includegraphics[width=0.9\linewidth,height=.6\linewidth]{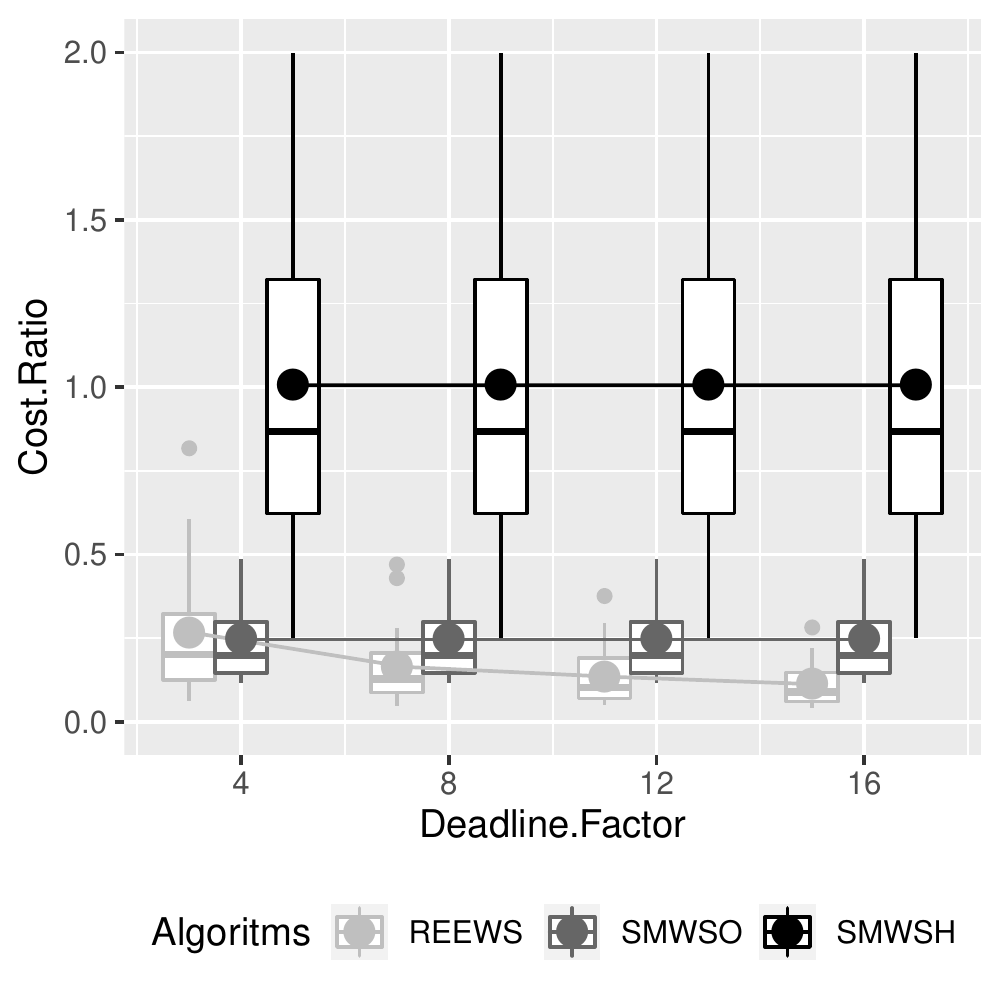}
			\caption{Cost Efficiency}\label{fig_montage_TimeRatio_and_CostRatio_a_SMWS_static}
		\end{subfigure}
		\begin{subfigure}[t]{.325\linewidth}
			\includegraphics[width=0.9\linewidth,height=.6\linewidth]{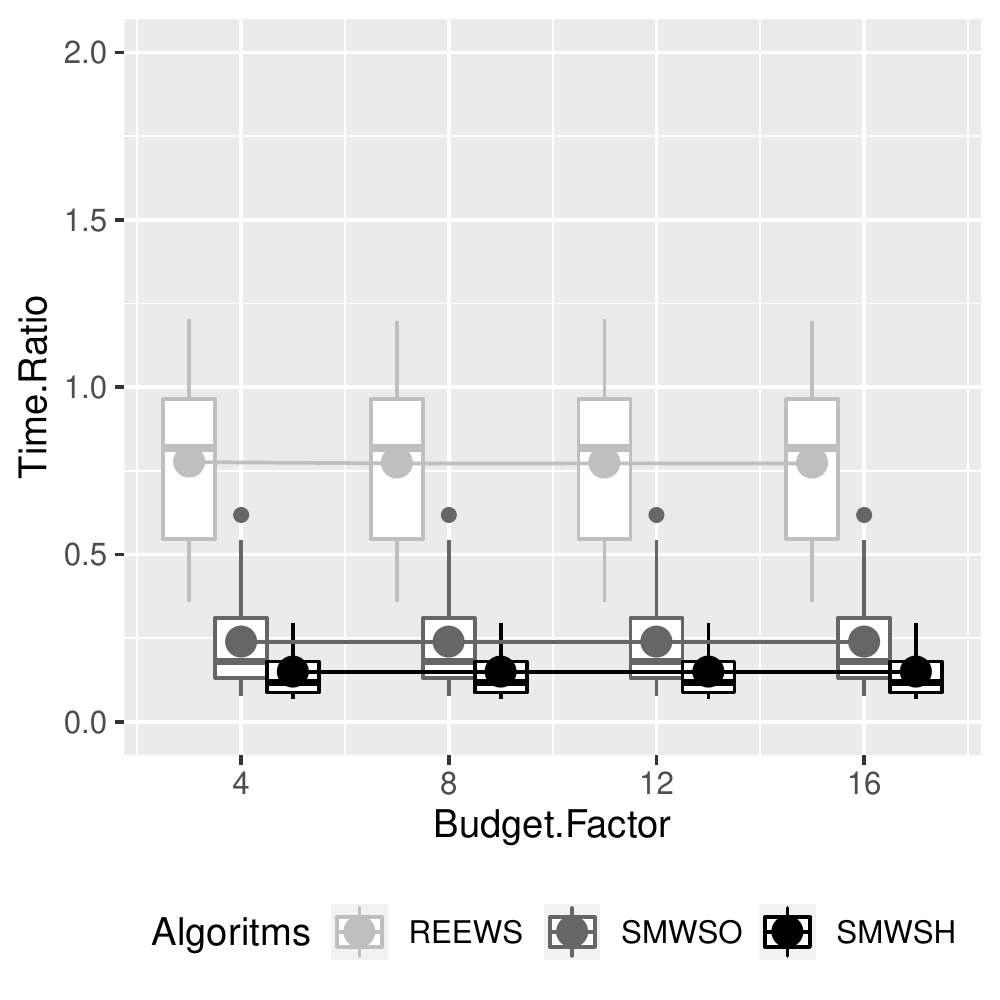}
			\caption{Time Efficiency}\label{fig_montage_TimeRatio_and_CostRatio_b_SMWS_static}
		\end{subfigure}		
		\begin{subfigure}[t]{.325\linewidth}
			\includegraphics[width=0.9\linewidth,height=.6\linewidth]{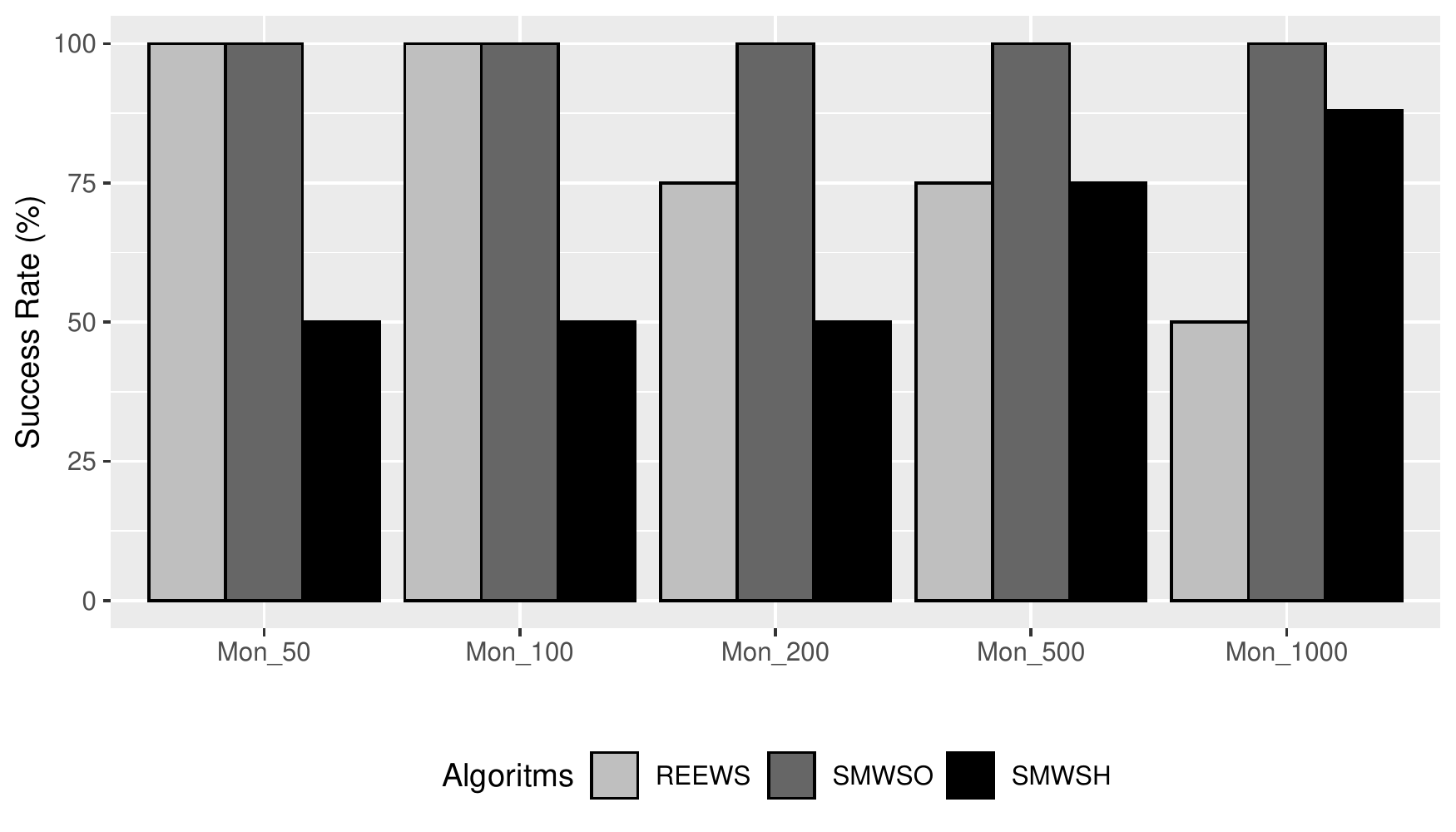}
			\caption{Success rate (\%)\label{fig_SuccessRate_Summary_MONTAGE_SMWS_static}}
		\end{subfigure}
		\begin{subfigure}[t]{.45\linewidth}
			\includegraphics[width=0.9\linewidth,height=.5\linewidth]{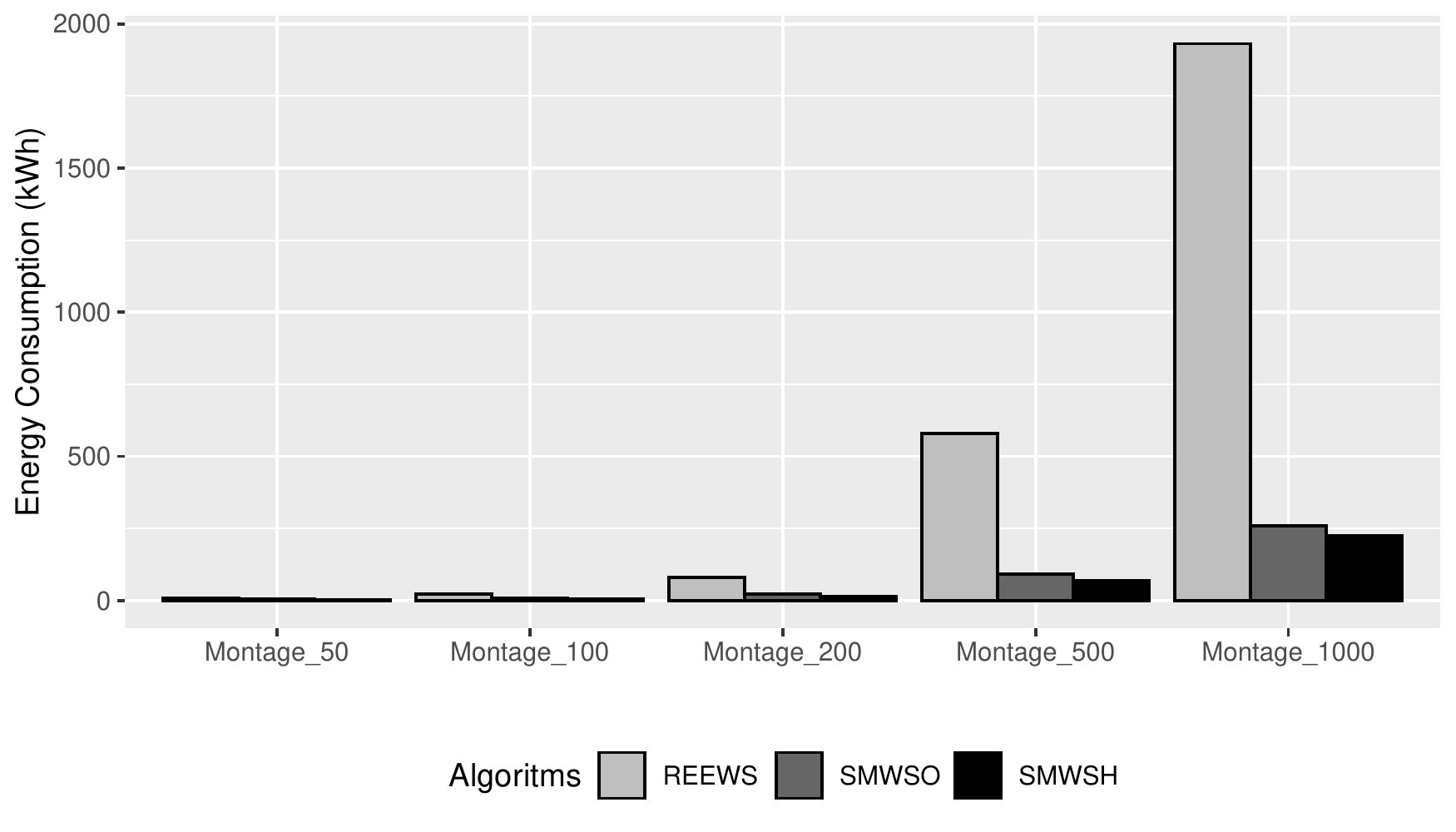}
			\caption{Energy consumption}	\label{fig_EnergyConsumption_MONTAGE_SMWS_static}		
		\end{subfigure}	
		
		\caption{Cost efficiency, time efficiency, and success rate (\%), Energy consumption of SMWSO and SMWSH vs REEWS for MONTAGE workflow.}\label{fig_montage_TimeRatio_and_CostRatio_and_SR_and_EnergyConsumption_SMWS_static} 
		
	\end{figure*}	
	
	\begin{itemize}	
		\item Cost Ratio (CR): The CR metric is used to compare the achieved costs of scheduling algorithms. A CR less than 1 means a schedule under the budget (see Eq. \ref{eq_CR}).
		\begin{equation}
			\label{eq_CR}
			CR = \dfrac{Cost_G}{B};
		\end{equation}
		
		\item Time Ratio (TR): In a similar way, the TR metric is used to compare the achieved times of scheduling algorithms. A TR less than 1 means a schedule under the deadline (see Eq. \ref{eq_TR}).
		\begin{equation}
			\label{eq_TR}
			TR = \dfrac{M_G}{\delta};
		\end{equation}
		
		\item Success Rate (SR): The SR metric is used to compare the rate of successes of the algorithms (see Eq. \ref{eq_SR}). There is success when both the CR and the TR are less than 1.
		\begin{equation}
			\label{eq_SR}
			SR = \dfrac{NB_{success}}{NB_{Exp}};
		\end{equation}
		
		$NB_{Exp}$ is total number of experiments, and $NB_{success}$ the number of successful schedules.
		
		\item Energy consumption: The energy in kilowatt-hours (kWh) consumed by the used VMs during the observed time.
	\end{itemize}

	\section{Simulation results and Analysis}
	\label{subsec:ProposedMOAlgo_Results_and_Analysis}
	
	in this section we present and analyze the results of the simulation. We first analyze the results for each of the five scientific workflows used in our experiments. Afterwards, we propose a summarized analysis of the results.
	
	The results are presented via diverse graphs, which show the performance of the different algorithms in terms of CR, TR, SR, and energy consumption. However, in order to do objectives analysis of the results of have conducted statistical tests (ANOVA with Tukey-Kramer post hoc test). Since the energy consumption is highly dependent from the workflow type and from the workfload, the statistical tests have been conducted by workflow and workload. During the simulation, they were a variation of four budget factors (4, 8, 12, 16) and four deadline factors (4, 8, 12, 16). Therefore, for each workload of each workflow, we have 16 different experiments. The summarized statistical tests are given in section \ref{subsec:ProposedMOAlgo_PerformanceSummary} for both  SR and energy efficiency.	
	
	\subsection{Performance for MONTAGE workflow}
	
	Figure \ref{fig_montage_TimeRatio_and_CostRatio_and_SR_and_EnergyConsumption_SMWS_static} presents the results obtained for MONTAGE workflow by REEWS, SMWSO and SMWSH. 
	In terms of time efficiency (see Fig. \ref{fig_montage_TimeRatio_and_CostRatio_b_SMWS_static}), while our two algorithms, SMWSO and SMWSH always have 100\% of schedules in the deadline, REEWS has less than 75\% of schedules in the deadline. However, In terms of cost efficiency (see Fig. \ref{fig_montage_TimeRatio_and_CostRatio_a_SMWS_static}), SMWSH has about 60\% of schedules in the budget while SMWSO and REEWS have 100\% of schedules in the budget.
	
	In terms of average success rate, while SMWSO realized 100\%, REEWS and SMWSH recorded respectively 80.00\% and 62.50\% (see Fig. \ref{fig_SuccessRate_Summary_MONTAGE_SMWS_static}).

	We noticed a significant influence of the workload of MONTAGE workflow over REEWS. When the number of tasks increases, the performance of REEWS decreases. This influence is found in SMWSH, but in reverse. When the number of tasks increases, the performance of SMWSH also increases.
	
	\begin{figure*}[!t]
		\centering
		\begin{subfigure}[t]{.325\linewidth}
			\includegraphics[width=0.9\linewidth,height=0.6\linewidth]{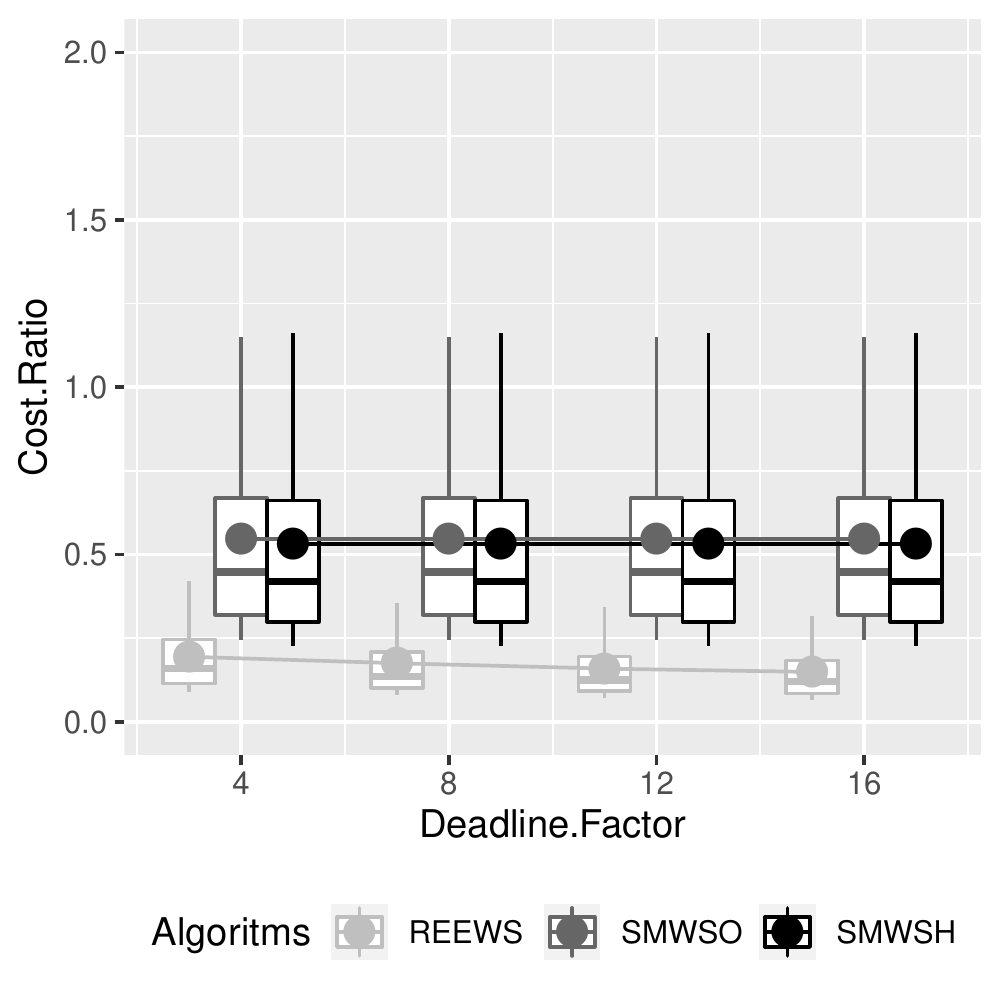}
			\caption{Cost Efficiency}\label{fig_cybershake_TimeRatio_and_CostRatio_a_SMWS_static}
		\end{subfigure}
		\begin{subfigure}[t]{.325\linewidth}
			\includegraphics[width=0.9\linewidth,height=0.6\linewidth]{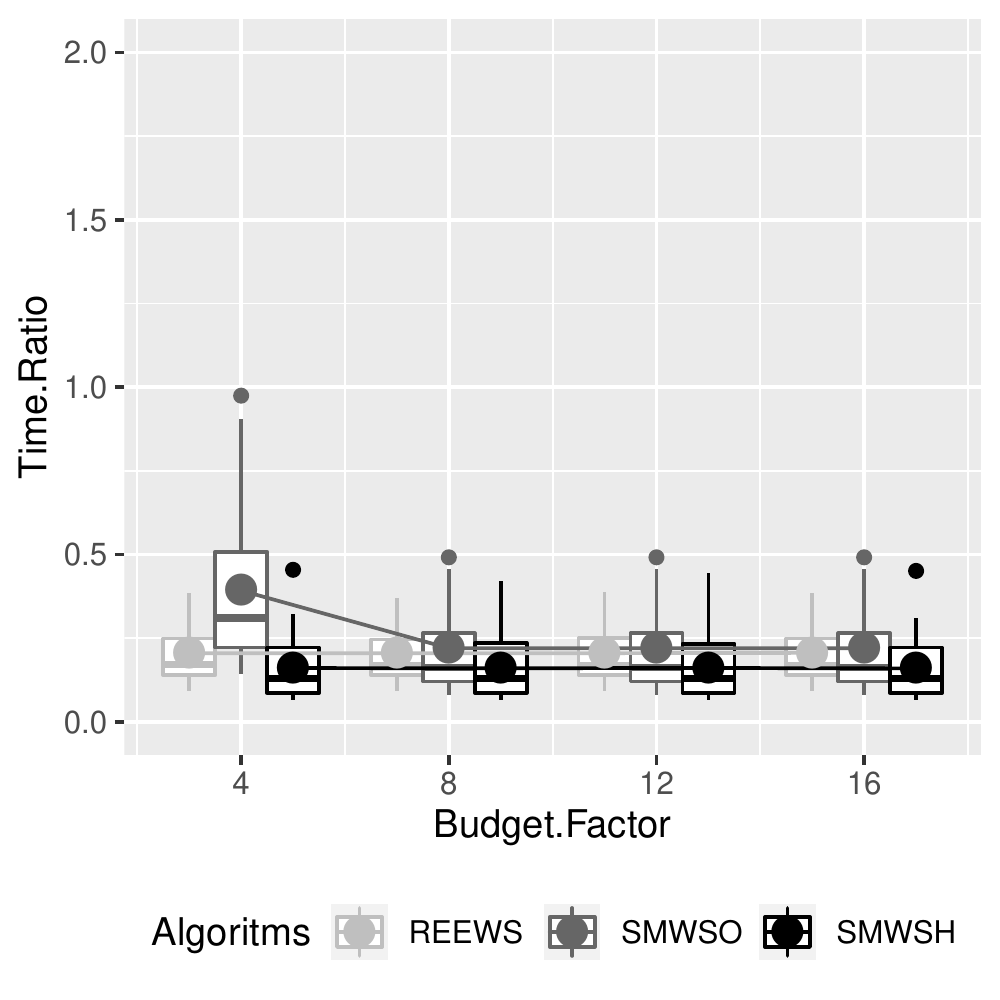}
			\caption{Time Efficiency}\label{fig_cybershake_TimeRatio_and_CostRatio_b_SMWS_static}
		\end{subfigure}		
		\begin{subfigure}[t]{.325\linewidth}
			\includegraphics[width=0.9\linewidth,height=0.6\linewidth]{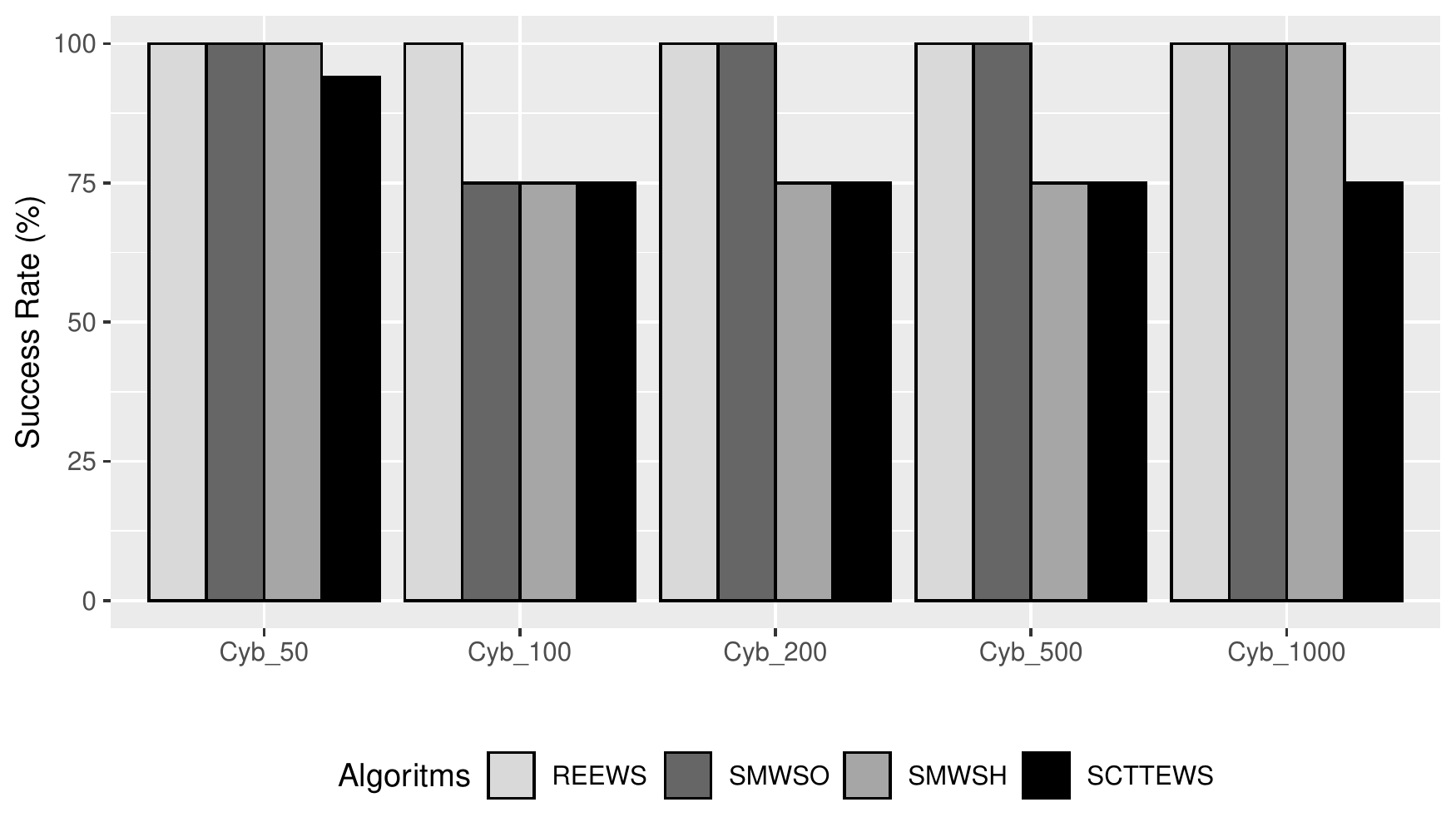}
			\caption{Success rate (\%)\label{fig_SuccessRate_Summary_CYBERSHAKE_SMWS_static}}
		\end{subfigure}
		\begin{subfigure}[t]{.45\linewidth}
			\includegraphics[width=0.9\linewidth,height=0.4\linewidth]{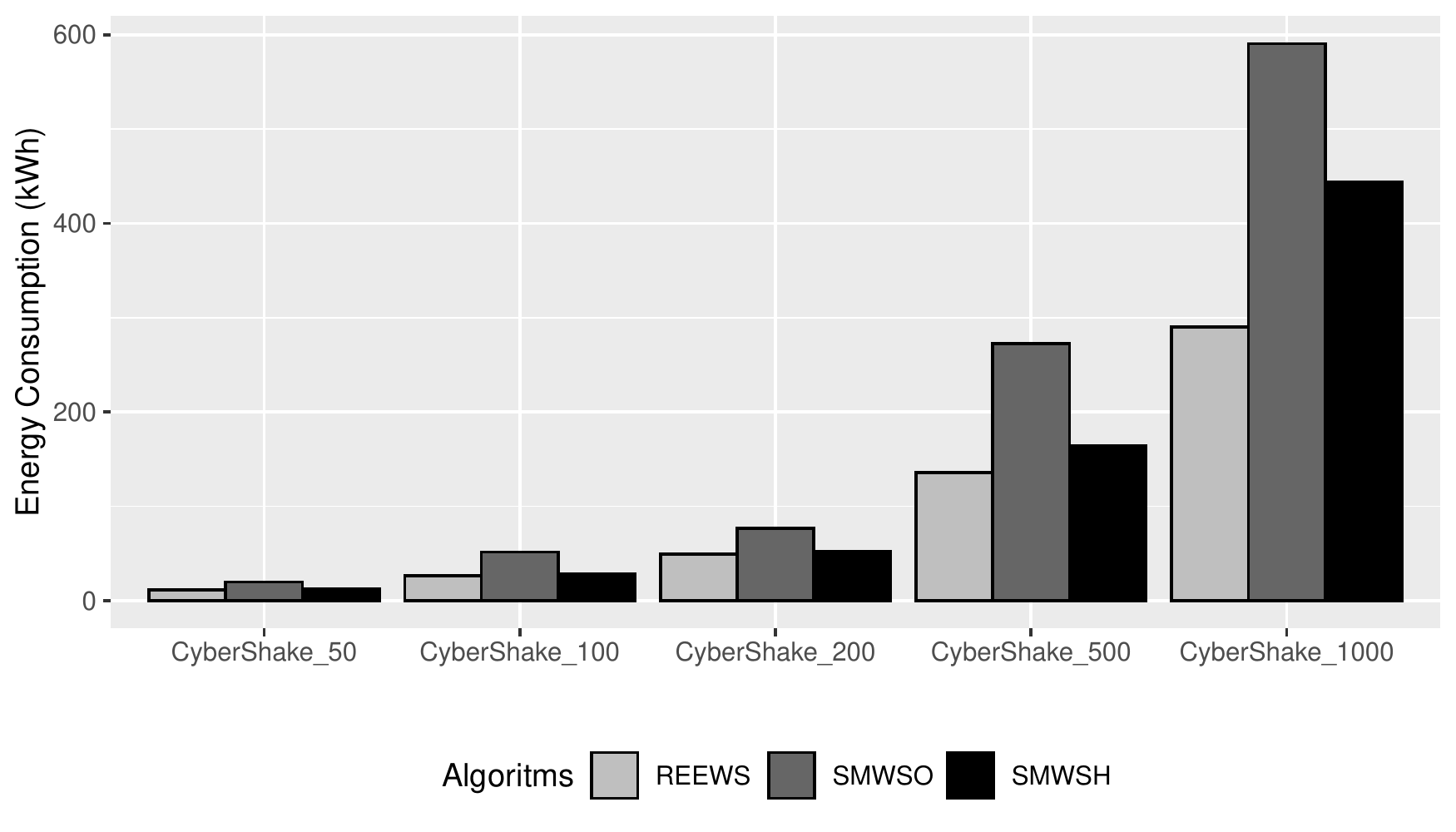}
			\caption{Energy consumption}	\label{fig_EnergyConsumption_CYBERSHAKE_SMWS_static}		
		\end{subfigure}	
		
		\caption{Cost efficiency, time efficiency, and success rate (\%), Energy consumption of SMWSO and SMWSH vs REEWS for CYBERSHAKE workflow.}\label{fig_cybershake_TimeRatio_and_CostRatio_and_SR_and_EnergyConsumption_SMWS_static} 
		
	\end{figure*}
	
	\begin{figure*}[!t]
		\centering
		\begin{subfigure}[t]{.325\linewidth}
			\includegraphics[width=0.9\linewidth,height=0.6\linewidth]{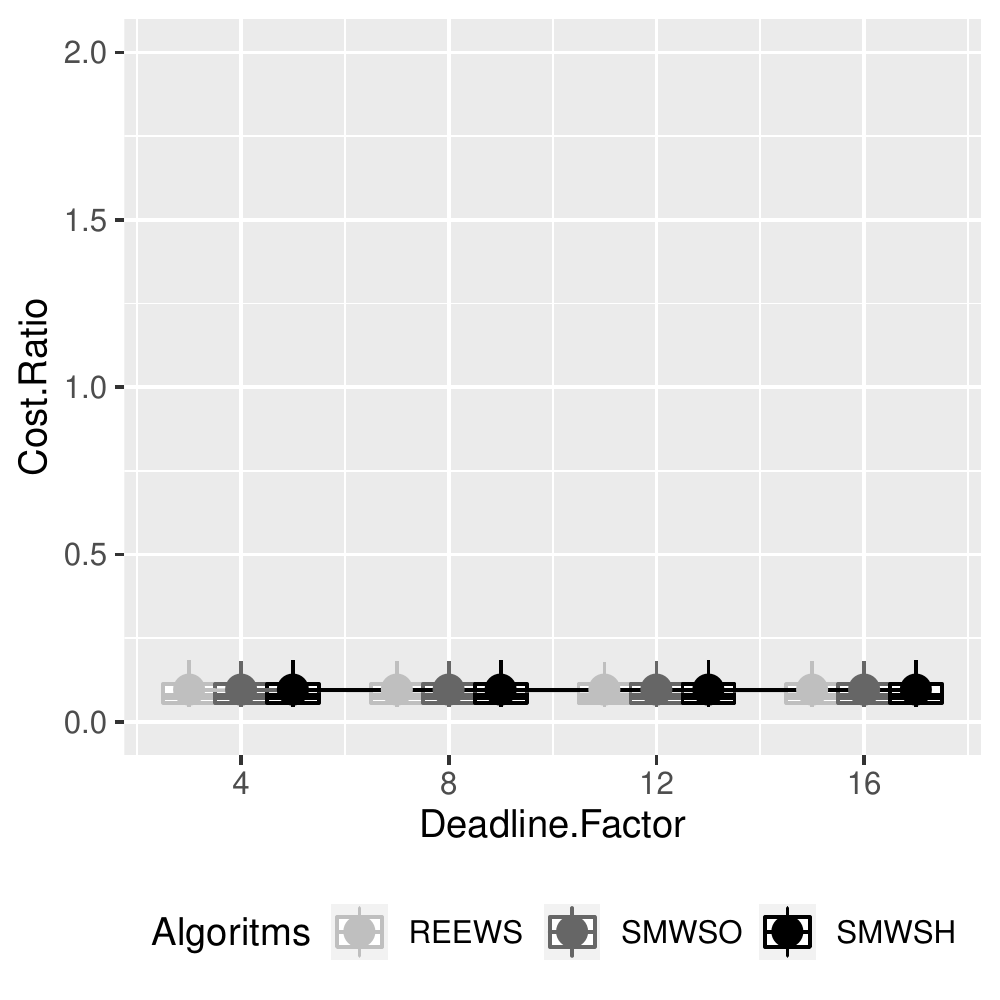}
			\caption{Cost Efficiency}\label{fig_epigenomics_TimeRatio_and_CostRatio_a_SMWS_static}
		\end{subfigure}
		\begin{subfigure}[t]{.325\linewidth}
			\centering\includegraphics[width=0.9\linewidth,height=0.6\linewidth]{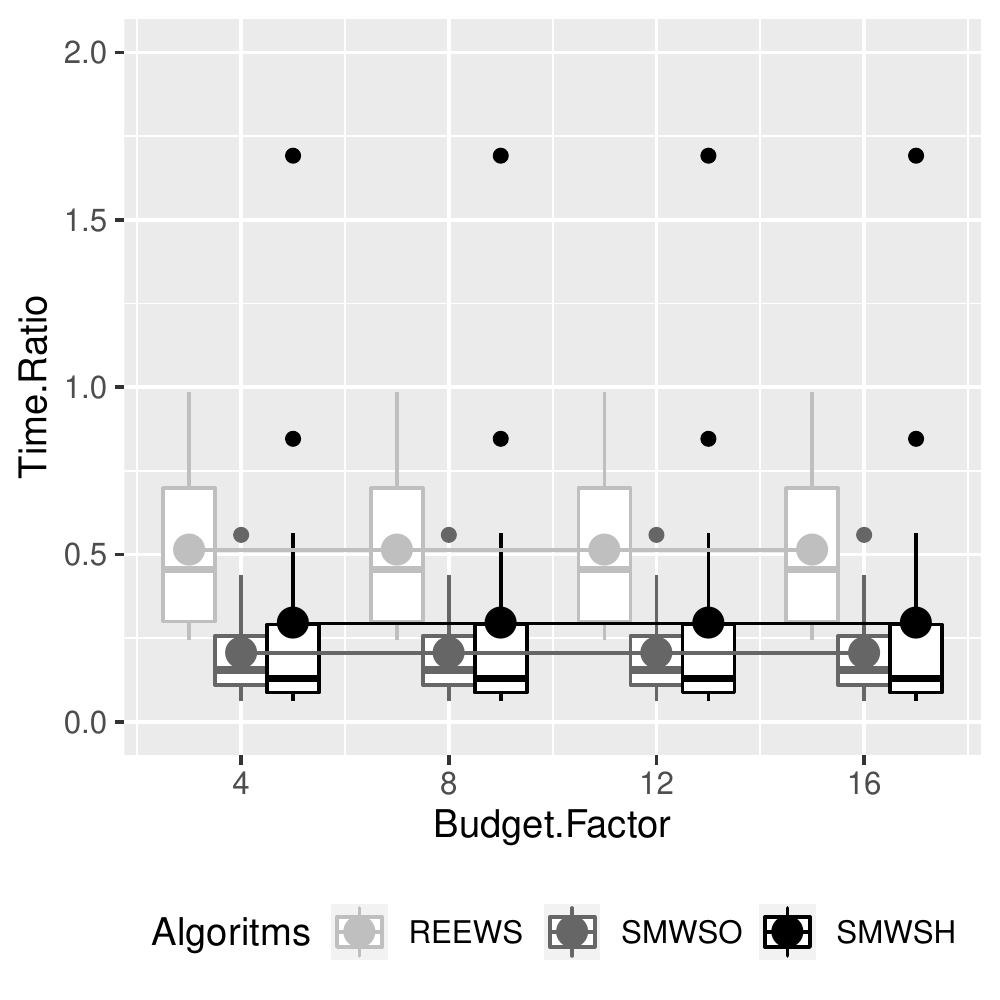}
			\caption{Time Efficiency}\label{fig_epigenomics_TimeRatio_and_CostRatio_b_SMWS_static}
		\end{subfigure}		
		\begin{subfigure}[t]{.325\linewidth}
			\includegraphics[width=.9\linewidth,height=0.6\linewidth]{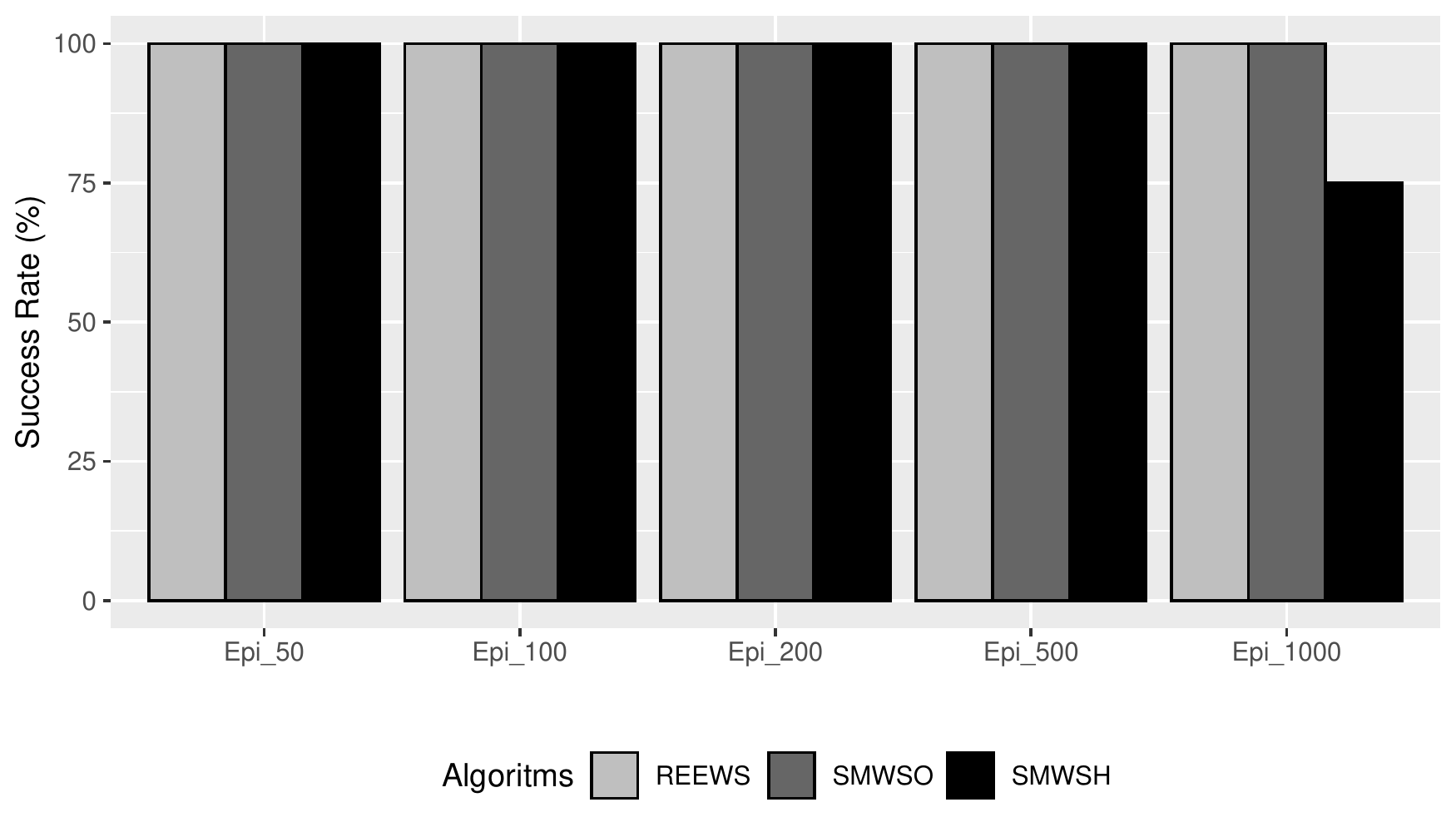}
			\caption{Success rate (\%)\label{fig_SuccessRate_Summary_EPIGENOMICS_SMWS_static}}
		\end{subfigure}
		\begin{subfigure}[t]{.45\linewidth}
			\includegraphics[width=0.9\linewidth,height=0.4\linewidth]{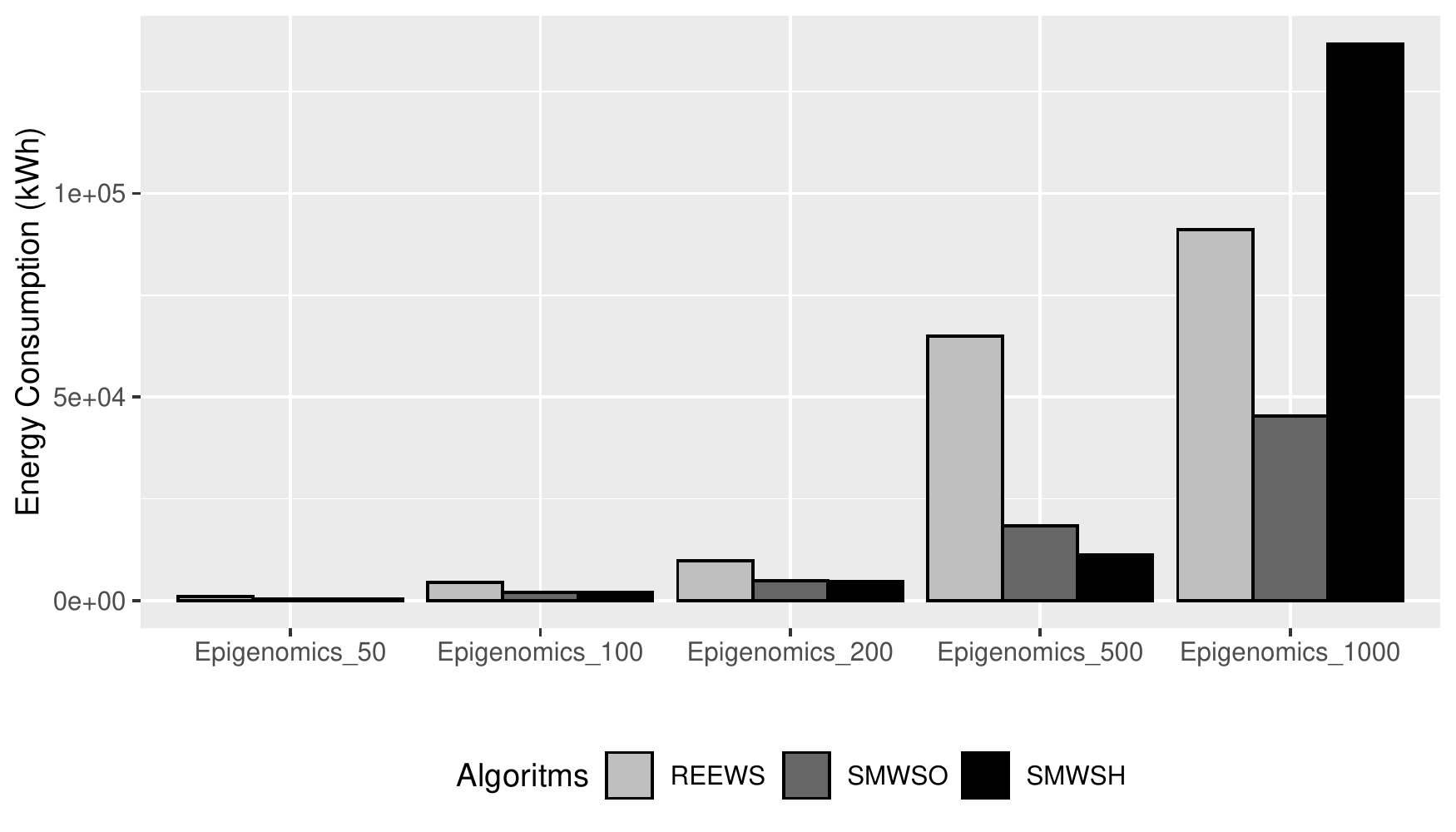}
			\caption{Energy consumption\label{fig_EnergyConsumption_EPIGENOMICS_SMWS_static}}			
		\end{subfigure}	
		
		\caption{Cost efficiency, time efficiency, and success rate (\%), Energy consumption of SMWSO and SMWSH vs REEWS for EPIGENOMICS workflow.}\label{fig_epigenomics_TimeRatio_and_CostRatio_and_SR_and_EnergyConsumption_SMWS_static} 
		
	\end{figure*}
	
	From Figure \ref{fig_EnergyConsumption_MONTAGE_SMWS_static}, we observe that the energy consumption of REEWS for MONTAGE workflow is greater than the ones of SMWSO, and SMWSH, as the number of tasks increases. That increase in energy consumption observed on REEWS is traceable to the increase of deadline missed due to the workload. In fact, Figure \ref{fig_SuccessRate_Summary_MONTAGE_SMWS_static} reveals a decrease of the success rate of REEWS due to the workload, and Figures \ref{fig_montage_TimeRatio_and_CostRatio_a_SMWS_static} and \ref{fig_montage_TimeRatio_and_CostRatio_b_SMWS_static} reveal that REEWS only fails because of deadline violation.

	\subsection{Performance for CYBERSHAKE workflow}

	For CYBERSHAKE workflow, SMWSO, SMWSH, and REEWS have a good time efficiency (see Fig. \ref{fig_cybershake_TimeRatio_and_CostRatio_b_SMWS_static}). However, in terms of cost efficiency only REEWS realizes 100\% of schedules in the budget, whereas SMWSO and SMWSH have 25\% of schedules out of the budget (see Fig. \ref{fig_cybershake_TimeRatio_and_CostRatio_b_SMWS_static}).
	
	\begin{figure*}[!t]
		\centering
		\begin{subfigure}[t]{.325\linewidth}
			\includegraphics[width=0.9\linewidth,height=0.6\linewidth]{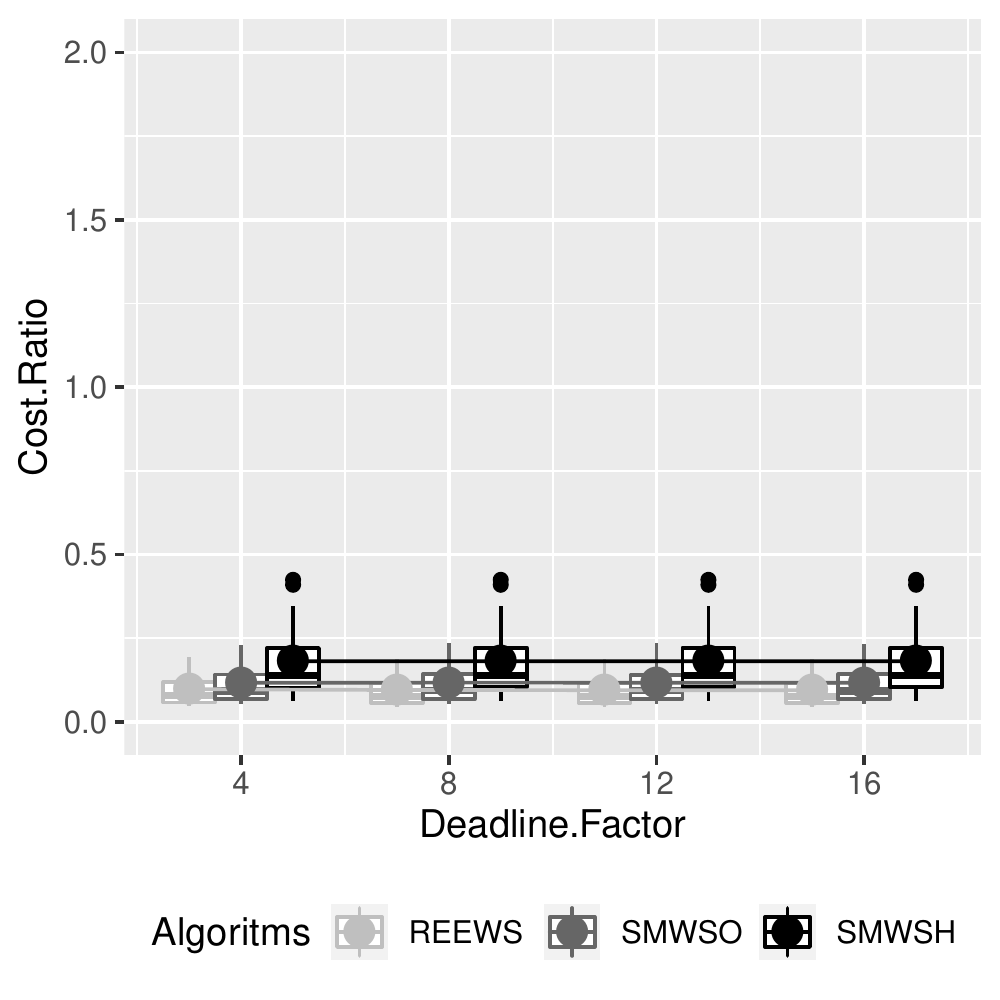}
			\caption{Cost Efficiency}\label{fig_sipht_TimeRatio_and_CostRatio_a_SMWS_static}
		\end{subfigure}
		\begin{subfigure}[t]{.325\linewidth}
			\includegraphics[width=0.9\linewidth,height=0.6\linewidth]{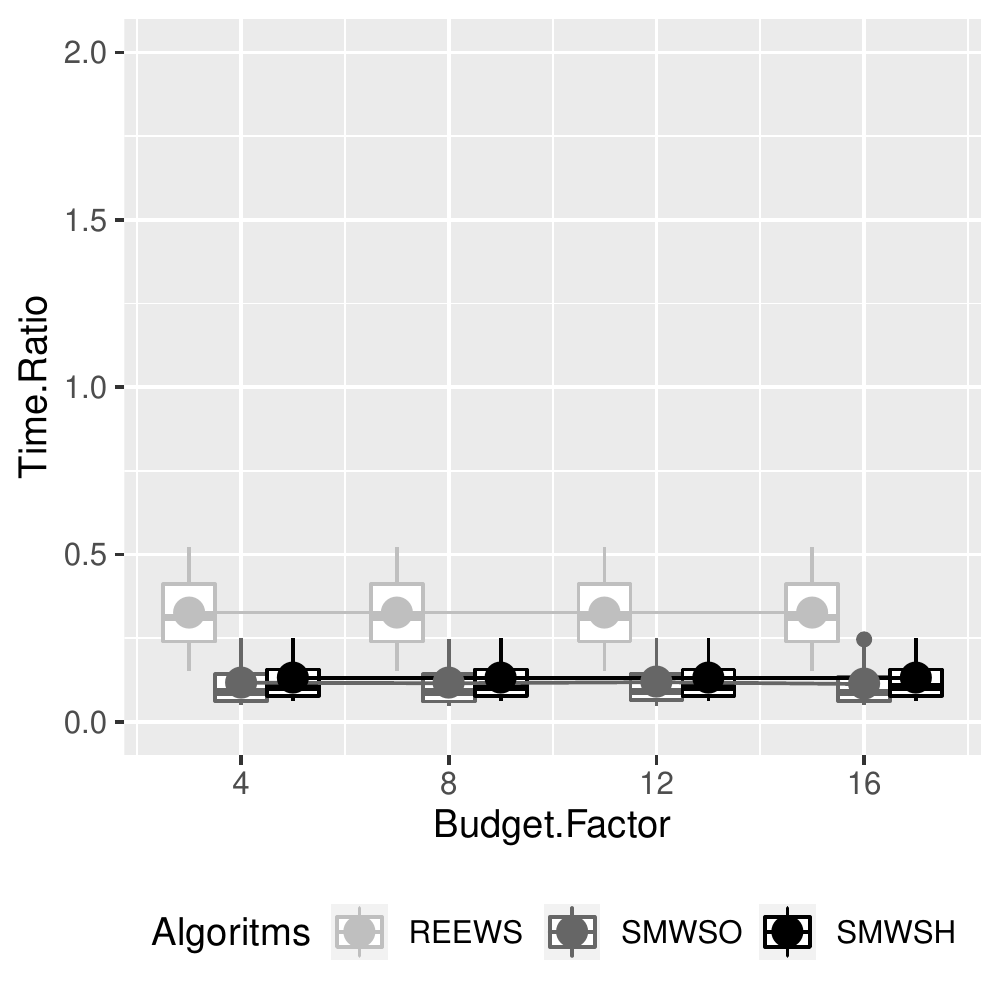}
			\caption{Time Efficiency}\label{fig_sipht_TimeRatio_and_CostRatio_b_SMWS_static}
		\end{subfigure}		
		\begin{subfigure}[t]{.325\linewidth}
			\includegraphics[width=0.9\linewidth,height=0.6\linewidth]{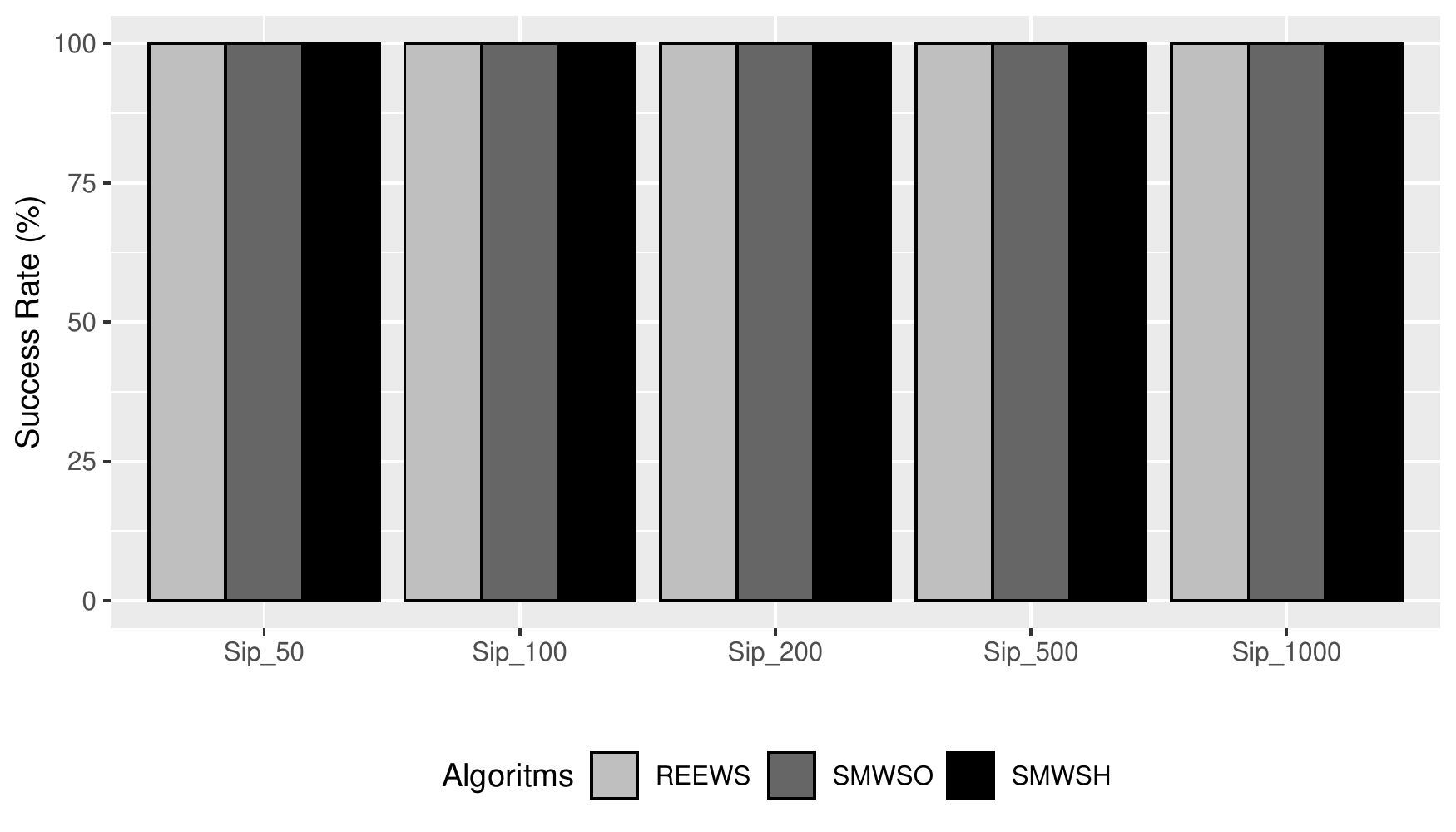}
			\caption{Success rate (\%)\label{fig_SuccessRate_Summary_SIPHT_SMWS_static}}	
		\end{subfigure}
		\begin{subfigure}[t]{.45\linewidth}
			\includegraphics[width=0.9\linewidth,height=0.4\linewidth]{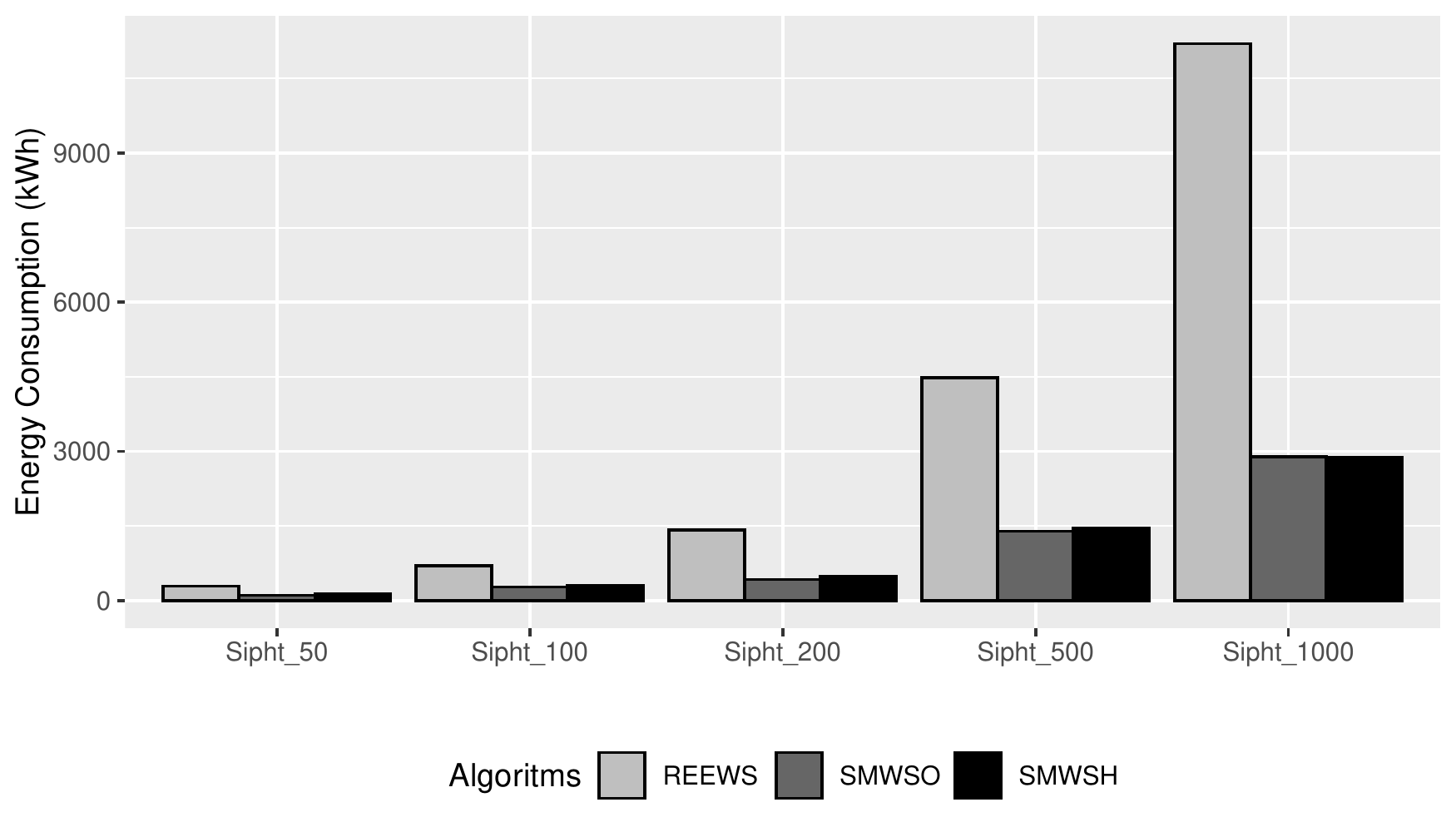}
			\caption{Energy consumption\label{fig_EnergyConsumption_SIPHT_SMWS_static}}		
		\end{subfigure}	
		
		\caption{Cost efficiency, time efficiency, and success rate (\%), Energy consumption of SMWSO and SMWSH vs REEWS for SIPHT workflow.}\label{fig_sipht_TimeRatio_and_CostRatio_and_SR_and_EnergyConsumption_SMWS_static} 
		
	\end{figure*}
	
	In terms of average success rate, REEWS recorded 100\% whereas SMWSO and SMWSH realized respectively 78.75\% and 85.00\%. REEWS is more energy-efficient than SMWSO, and SMWSH for CYBERSHAKE workflow (see Fig. \ref{fig_EnergyConsumption_CYBERSHAKE_SMWS_static}).

	\subsection{Performance for EPIGENOMICS workflow}
	
	For EPIGENOMICS workflow, SMWSO, SMWSH, and REEWS have 100\% of schedules in the budget (see Fig. \ref{fig_epigenomics_TimeRatio_and_CostRatio_a_SMWS_static}). However, in terms of cost efficiency only SMWSO and REEWS realise 100\% of schedules in the deadline, whereas SMWSH have few schedules out of the deadline (see Fig. \ref{fig_epigenomics_TimeRatio_and_CostRatio_b_SMWS_static}).
	
	In terms of average success rate, SMWSO and REEWS recorded 100\% whereas SMWSH realized 95.00\%.

	SMWSO is more energy-efficient than REEWS and SMWSH for EPIGENOMICS workflow (see Fig. \ref{fig_EnergyConsumption_EPIGENOMICS_SMWS_static}). We notice that for the workload of 1000 tasks, SMWSH recorded greater energy consumption compared to SMWSO and REEWS. This was not the case for the smaller workload.

	\subsection{Performance for SIPHT workflow}
	
	For SIPHT workflow, SMWSO, SMWSH, and REEWS have 100\% of schedules in both the budget and the deadline (see Fig. \ref{fig_sipht_TimeRatio_and_CostRatio_a_SMWS_static} and \ref{fig_sipht_TimeRatio_and_CostRatio_b_SMWS_static}). Therefore, SMWSO, SMWSH, and REEWS have 100\% of average success rate. 
	
	REEWS is less energy-efficient than SMWSO and SMWSH for SIPHT workflow (see Fig. \ref{fig_EnergyConsumption_SIPHT_SMWS_static}). The workload of SIPHT also has significant impact on the energy-efficiency of REEWS.

	\subsection{Performance for LIGO workflow}
	
\begin{figure*}[!t]
		\centering
		\begin{subfigure}[t]{.325\linewidth}
			\includegraphics[width=0.9\linewidth,height=0.6\linewidth]{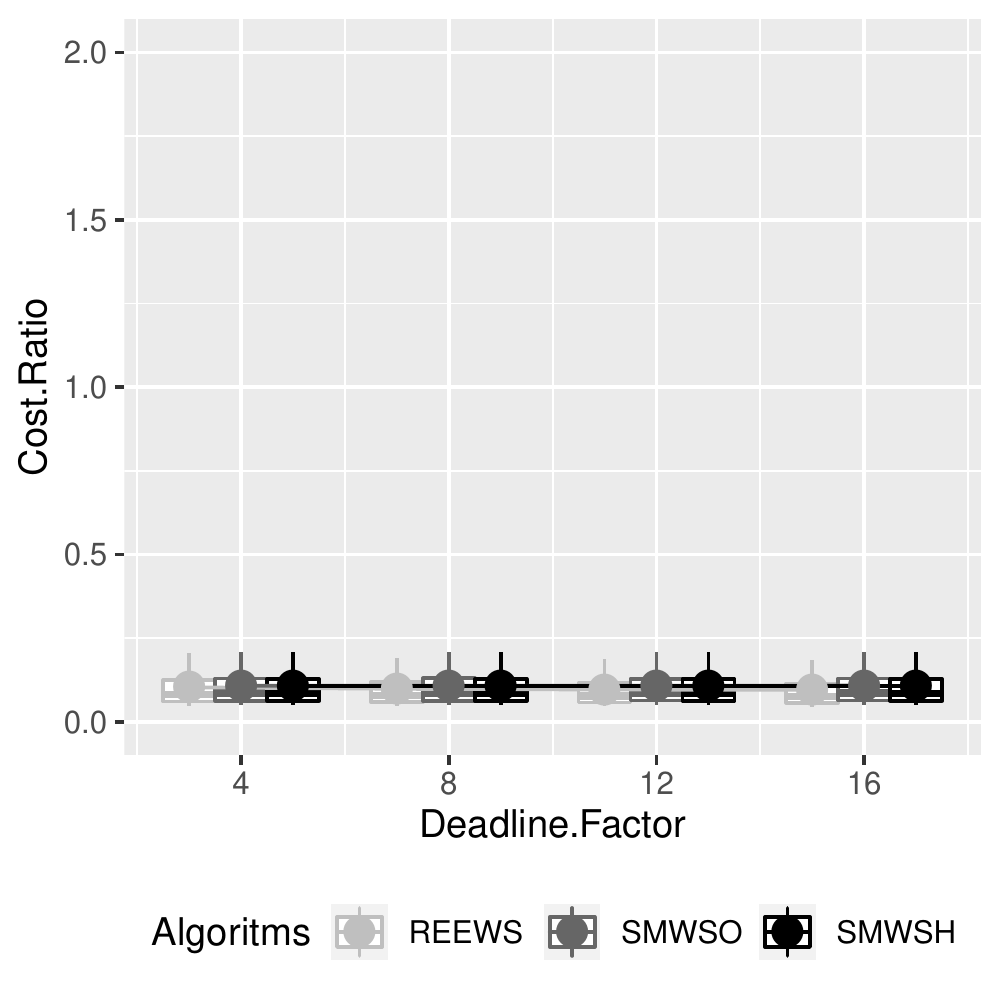}
			\caption{Cost Efficiency}\label{fig_ligo_TimeRatio_and_CostRatio_a_SMWS_static}
		\end{subfigure}
		\begin{subfigure}[t]{.325\linewidth}
			\includegraphics[width=0.9\linewidth,height=0.6\linewidth]{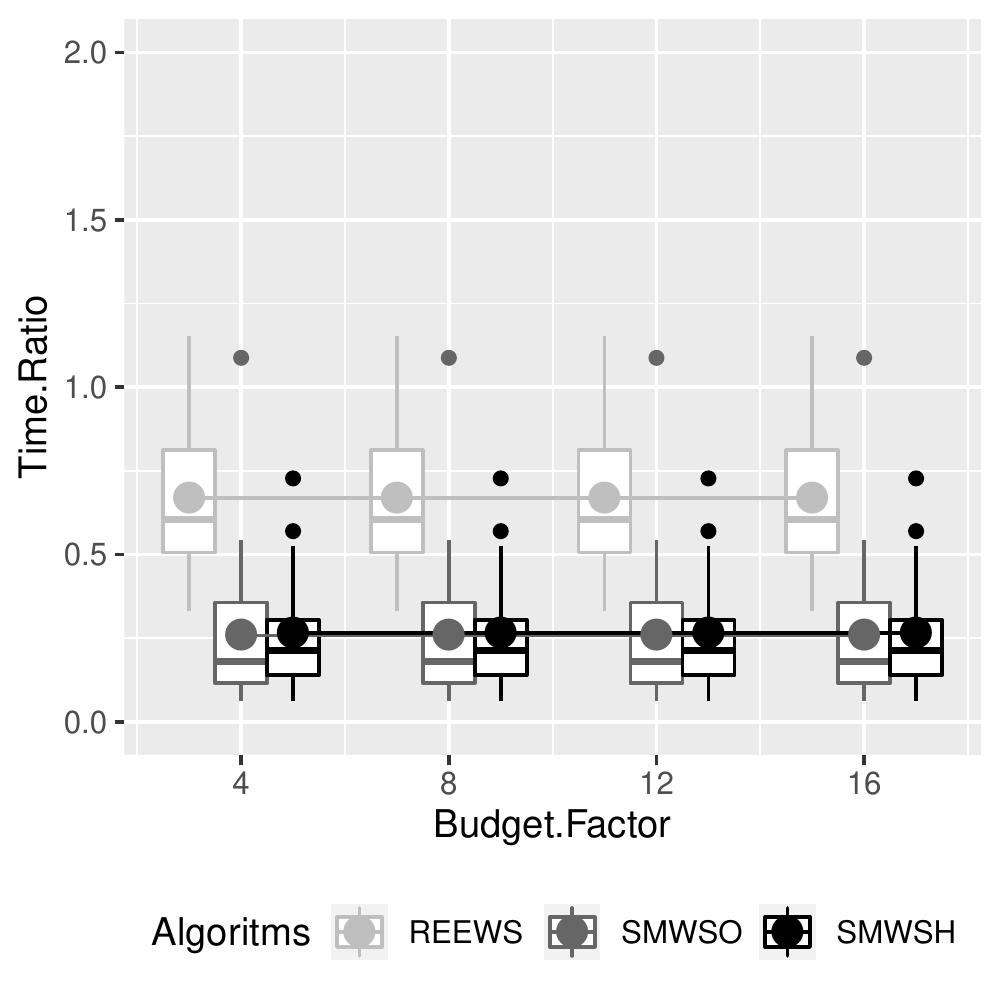}
			\caption{Time Efficiency}\label{fig_ligo_TimeRatio_and_CostRatio_b_SMWS_static}
		\end{subfigure}		
		\begin{subfigure}[t]{.325\linewidth}
			\includegraphics[width=0.9\linewidth,height=0.6\linewidth]{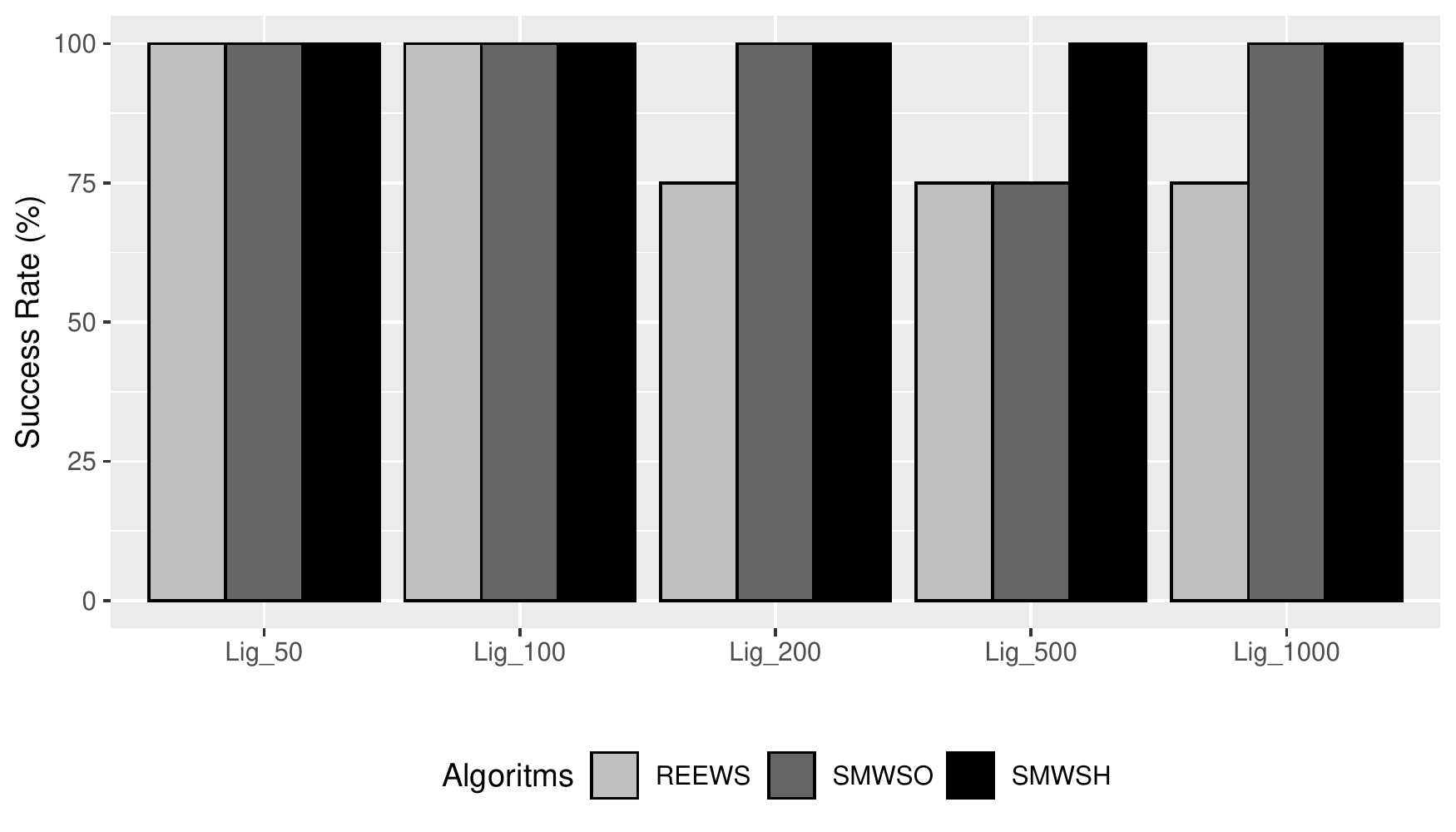}
			\caption{Success rate (\%)\label{fig_SuccessRate_Summary_LIGO_SMWS_static}}		
		\end{subfigure}
		\begin{subfigure}[t]{.45\linewidth}
			\includegraphics[width=0.9\linewidth,height=0.4\linewidth]{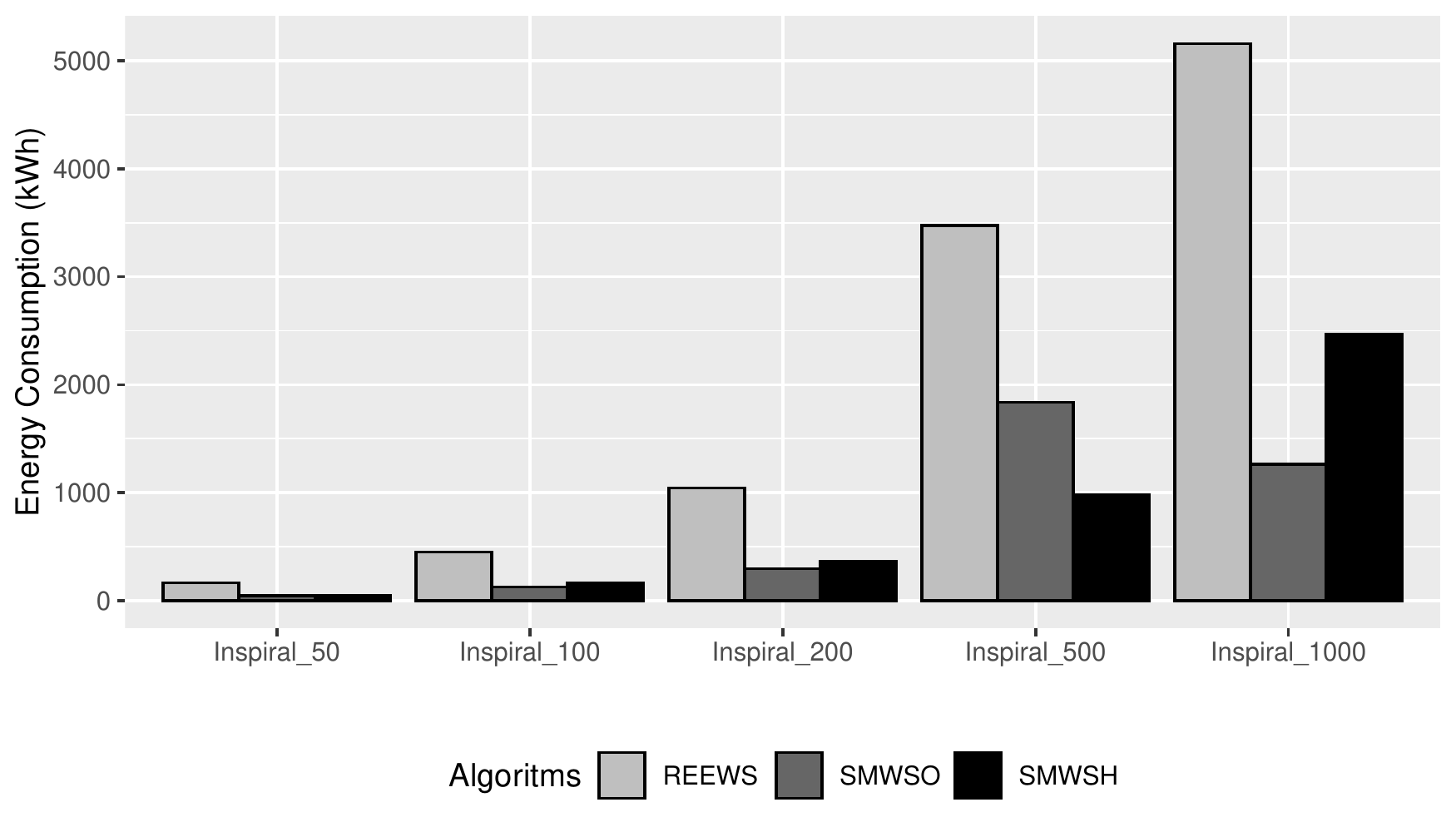}
			\caption{Energy consumption}\label{fig_EnergyConsumption_LIGO_SMWS_static}
		\end{subfigure}
		
		\caption{Cost efficiency, time efficiency, and success rate (\%), Energy consumption of SMWSO and SMWSH vs REEWS for LIGO workflow.}\label{fig_ligo_TimeRatio_and_CostRatio_and_SR_and_EnergyConsumption_SMWS_static} 
		
	\end{figure*}

	For LIGO workflow, SMWSH has 100\% of schedules in both the budget and the deadline (see Fig. \ref{fig_ligo_TimeRatio_and_CostRatio_a_SMWS_static} and \ref{fig_ligo_TimeRatio_and_CostRatio_b_SMWS_static}). SMWSO and REEWS have few schedules out of the deadline and 100\% of schedules in the budget. 
	
	The average of success rate of SMWSH is 100\%, whereas SMWSO and REEWS have of respectively 95.00\% and 85.00\%.
	
	REEWS is less energy-efficient than SMWSO and SMWSH for LIGO workflow (see Fig. \ref{fig_EnergyConsumption_LIGO_SMWS_static}). The workload of LIGO also has significant impact on the energy-efficiency of REEWS. 
	
	\subsection{Performance summary and discussions}
	\label{subsec:ProposedMOAlgo_PerformanceSummary} 
	
	In this subsection, we present a summary of the simulation results and provide some analysis.

	\subsubsection{In terms of success rate}

	In terms of Success Rate, the three algorithms REEWS, SMWSO, and SMWSH recorded respectively 93.00\%, \textbf{98.00}\%, and 88.50\% as mean of SR (see Table \ref{tab_ANOVA_Test_SR_of_REEWS__SMWSO_and_SMWSH__a}). SMWSO has higher success rate than the other, with the smallest standard deviation (\textit{2.44}). However, ANOVA test reveals that REEWS, SMWSO, and SMWSH have comparable performances in terms of success rate (see Table \ref{tab_ANOVA_Test_SR_of_REEWS__SMWSO_and_SMWSH__b}).

	\begin{table}[!b]
		\centering
		\caption{ANOVA test result comparing the SR of the three algorithms REEWS, SMWSO and SMWSH} \label{tab_ANOVA_Test_SR_of_REEWS__SMWSO_and_SMWSH}
		
		
		\begin{subtable}{0.7\linewidth}
			\centering
			\caption{Summary of input}
			\resizebox{\linewidth}{!}{\begin{tabular}{l c c c c }
					\hline
					\textit{\textbf{Group}} & \textit{\textbf{Count}} & \textit{\textbf{Sum}} & \textit{\textbf{Average}} & \textit{\textbf{Variance}} \\
					\hline
					\textit{\textbf{REEWS}} & 25 & 2325	&93	&183.33 \\
					
					\textit{SMWSO} 			  & 25 & 2450	&98	&47.92 \\
					
					\textit{SMWSH} 		  & 25 & 2213	&88.52	&311.43 \\
					\hline
					\\
			\end{tabular}}

			\label{tab_ANOVA_Test_SR_of_REEWS__SMWSO_and_SMWSH__a}
		\end{subtable}
		\\
		\begin{subtable}{\linewidth}
			\centering
			
			\caption{ANOVA test result}
			\resizebox{\linewidth}{!}{\begin{tabular}{ l c c c c c c }
					\hline
					\textit{\textbf{Source of Variation}} & \textit{\textbf{SS}} & \textit{\textbf{df}} & \textit{\textbf{MS}} & \textit{\textbf{F stat.}} & \textit{\textbf{P-value}} & \textit{\textbf{F critical}}\\
					\hline
					\textit{\textbf{Between Groups}}  & 1124.51	&2	&562.25	& 3.11	&0.05	&3.12  \\
					
					\textit{Within Groups} 			  & 13024.24	&72	&180.90 &  &  &  \\
					
					\hline	
					\textit{Total} 			  		  & 14148.75 & 74 &  &  &  &  \\
					\hline
			\end{tabular}}

			\label{tab_ANOVA_Test_SR_of_REEWS__SMWSO_and_SMWSH__b}
		\end{subtable}
	\end{table}


	\subsubsection{In terms of energy efficiency}

	Since the type and the workload of the workflow highly influence energy consumption, some statistical tests have been conducted by workflow and workload using ANOVA with Tukey-Kramer post hoc tests. The tests results reveal a significant different between the energy consumption produced by the three algorithms. 
	
	Table \ref{tab_Energy_Efficiency_Ranking_Summary_for_MO_Static_Algorithms} presents the summary of the energy efficiency ranking between the three algorithms REEWS, SMWSO and SMWSH, obtained from statistical tests (you can further explore the determination of the ranking for the case of MONTAGE on Appendix \ref{appA:energy}). It can then be inferred that the REEWS is more energy-efficient or as energy-efficient as our proposals only for CYBERSHAKE (50, 100, 200, 500, and 1000). In all the other cases (80\% of cases), our two algorithms SMWSO and SMWSH are significantly more energy-efficient than REEWS.

\begin{figure}[!t]
		\centering\includegraphics[width=0.9\linewidth]{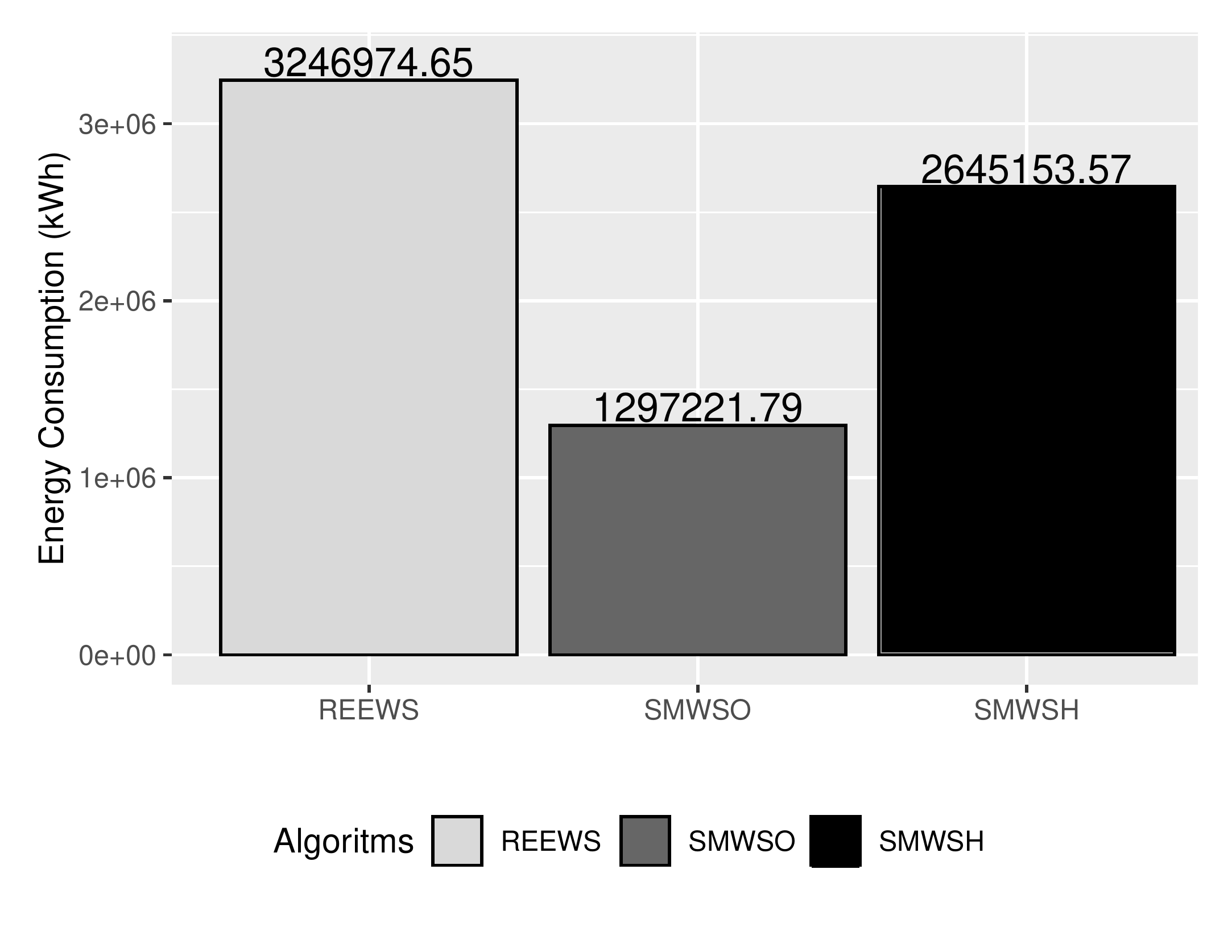}
		\vspace{-0.5cm}
		\caption{Total Energy consumption of SMWSO and SMWSH vs REEWS, for all the experiments.\label{fig_EnergyConsumption_total_SMWS_static}}
		\vspace{-0.4cm}\end{figure}

	The results prove the significant improvement of our proposal, the SMWSO algorithm, in terms of energy-saving. We have advocated that:
	
	\begin{itemize}
		
		
		\item \textit{Homogeneity can produce better results if good instances are chosen for the execution of the workflow.} The out-performance of SMWSO against SMWSH confirm our statement. In fact, the two algorithms are designed almost in the same manner, apart from the variety of resources used. Both algorithm are sometime more energy-efficient than the other one, almost equitably (see Table \ref{tab_Energy_Efficiency_Ranking_Summary_for_MO_Static_Algorithms}). However, it is more significant when SMWSO is the best, with a total energy-saving more than 50\% better (see Fig. \ref{fig_EnergyConsumption_total_SMWS_static}).
		
		
		\item \textit{If a suitable number of VMs determined, it can help not only to produce better results (in terms of success rate) but also upgrade the VM utilization Maximization and the Energy Consumption Minimization as well as the Workload Maximization.} In our two multi-objective algorithms, the optimal number of VMs technique is proposed as an answer to that preoccupation and used to limit the number of VMs to use. The result show that our two algorithms significantly outperform REEWS in terms of energy-saving in most of the types of workflow (apart of Cybershake) and of the workloads. 
		
	\end{itemize}
	\begin{table}[!t]
		\centering
		\caption{Energy efficiency ranking between the three algorithms REEWS, SMWSO and SMWSH, proceeded from ANOVA with Tukey-Kramer post hoc statistical tests.} 
		\label{tab_Energy_Efficiency_Ranking_Summary_for_MO_Static_Algorithms}
		\scriptsize
		\resizebox{\linewidth}{!}{\begin{tabular}{ c c c c }
				\hline
				\textbf{Workflow} & \textbf{REEWS}  & \textbf{SMWSO} & \textbf{SMWSH} \\
				\hline	
				MONTAGE 50 &3	&2	&1\\
				MONTAGE 100 &3	&2	&1\\
				MONTAGE 200 &3	&1	&1\\
				MONTAGE 500 &3	&1	&1\\
				MONTAGE 1000 &3	&1	&1\\

				\hline
				
				CYBERSHAKE 50 &1	&3	&2\\
				CYBERSHAKE 100 &1	&3	&1\\
				CYBERSHAKE 200 &1	&3	&1\\
				CYBERSHAKE 500 &1	&3	&1\\
				CYBERSHAKE 1000 &1	&3	&2\\
				
				\hline
				
				EPIGENOMICS 50 &3	&2	&1\\
				EPIGENOMICS 100 &3	&1	&1\\
				EPIGENOMICS 200 &3	&1	&1\\
				EPIGENOMICS 500 &3	&1	&1\\
				EPIGENOMICS 1000 &3	&1	&2\\
				
				\hline
				
				SIPHT 50 &3	&1	&2\\
				SIPHT 100 &3	&1	&1\\
				SIPHT 200 &3	&1	&1\\
				SIPHT 500 &3	&1	&1\\
				SIPHT 1000 &3	&1	&1\\
				
				\hline
				
				LIGO 50 &3	&1	&1\\ 
				LIGO 100 &3	&1	&1\\ 
				LIGO 200 &3	&1	&1\\ 
				LIGO 500 &3	&2	&1\\ 
				LIGO 1000 &3	&1	&2\\ 
				
				\hline
		\end{tabular}}
	\end{table}
	
	\section{Conclusion And Future Work}
	
	In this paper, we present the Structure-based Multi-objective Workflow Scheduling with an Optimal instance type (SMWSO) algorithm, aiming at optimizing processing costs, makespan, and energy consumption, under the user-defined budget and deadline in the cloud. SMWSO introduces a new concept, the optimal instance type determination along with the optimal number of VMs, and only uses VMs of optimal instance type. These techniques aim at avoiding resource wastage by determining the optimal type and limiting the number of VMs to provision to an optimal threshold. The employment of the DVFS on non-critical paths have been used to further reduce the energy consumption. Its heterogeneous VMs version denoted as SMWSH has been designed in order to highlight its strength within the heterogeneous environment.

	Comparative experimentation have been done through simulations against the state-of-the-art algorithm REEWS. Performance results supported by appropriate statistical tests prove the out-performance of our proposals in terms of energy-saving. SMWSO and SMWSH are more energy-efficient than REEWS in 80\% of cases (workflow / workload). Moreover, SMWSO is the more energy-saving, and can save more than 50\% total energy compared to SMWSH and REEWS. In terms of user satisfaction, while SMWSO scored at overall the highest success rate, statistical tests proved that their is no significant difference between the three algorithms. This confirms our hypothesis. Firstly, homogeneity can produce better results if a good instance is chosen for the execution of the workflow. Secondly, if a suitable number of VMs is determined, it can help to produce better results.
	
	In our future work, we intend to investigate on the scheduling of multiple concurrent workflows, and also to deal with the uncertainty nature of cloud environments by using machine learning.

	%
	\section*{Conflict of interest}	
	\noindent The authors declare that they have no conflict of interest.

	\nocite{*} 

\begin{thebibliography}{10}
		
		\bibitem{beloglazov2011taxonomy}
		Anton Beloglazov, Rajkumar Buyya, Young~Choon Lee, and Albert Zomaya.
		\newblock A taxonomy and survey of energy-efficient data centers and cloud
		computing systems.
		\newblock In {\em Advances in computers}, volume~82, pages 47--111. Elsevier,
		2011.
		
		\bibitem{duy2010performance}
		Truong Vinh~Truong Duy, Yukinori Sato, and Yasushi Inoguchi.
		\newblock Performance evaluation of a green scheduling algorithm for energy
		savings in cloud computing.
		\newblock In {\em 2010 IEEE international symposium on parallel \& distributed
			processing, workshops and Phd forum (IPDPSW)}, pages 1--8. IEEE, 2010.
		
		\bibitem{greenberg2008cost}
		Albert Greenberg, James Hamilton, David~A Maltz, and Parveen Patel.
		\newblock The cost of a cloud: research problems in data center networks, 2008.
		
		\bibitem{buyya2013introduction}
		Rajkumar Buyya.
		\newblock Introduction to the ieee transactions on cloud computing.
		\newblock {\em IEEE Transactions on cloud computing}, 1(1):3--21, 2013.
		
		\bibitem{reiss2012heterogeneity}
		Charles Reiss, Alexey Tumanov, Gregory~R Ganger, Randy~H Katz, and Michael~A
		Kozuch.
		\newblock Heterogeneity and dynamicity of clouds at scale: Google trace
		analysis.
		\newblock In {\em Proceedings of the Third ACM Symposium on Cloud Computing},
		pages 1--13, 2012.
		
		\bibitem{piraghaj2017survey}
		Sareh~Fotuhi Piraghaj, Amir~Vahid Dastjerdi, Rodrigo~N Calheiros, and Rajkumar
		Buyya.
		\newblock A survey and taxonomy of energy efficient resource management
		techniques in platform as a service cloud.
		\newblock In {\em Handbook of Research on End-to-End Cloud Computing
			Architecture Design}, pages 410--454. IGI Global, 2017.
		
		\bibitem{garg2020energy}
		Neha Garg, Damanpreet Singh, and Major~Singh Goraya.
		\newblock Energy and resource efficient workflow scheduling in a virtualized
		cloud environment.
		\newblock {\em Cluster Computing}.
		
		\bibitem{dabbagh2015toward}
		Mehiar Dabbagh, Bechir Hamdaoui, Mohsen Guizani, and Ammar Rayes.
		\newblock Toward energy-efficient cloud computing: Prediction, consolidation,
		and overcommitment.
		\newblock {\em IEEE network}, 29(2):56--61, 2015.
		
		\bibitem{beloglazov2012energy}
		Anton Beloglazov, Jemal Abawajy, and Rajkumar Buyya.
		\newblock Energy-aware resource allocation heuristics for efficient management
		of data centers for cloud computing.
		\newblock {\em Future generation computer systems}, 28(5):755--768, 2012.
		
		\bibitem{garg2019reliability}
		Ritu Garg, Mamta Mittal, et~al.
		\newblock Reliability and energy efficient workflow scheduling in cloud
		environment.
		\newblock {\em Cluster Computing}, 22(4):1283--1297, 2019.
		
		\bibitem{li2015cost}
		Zhongjin Li, Jidong Ge, Haiyang Hu, Wei Song, Hao Hu, and Bin Luo.
		\newblock Cost and energy aware scheduling algorithm for scientific workflows
		with deadline constraint in clouds.
		\newblock {\em IEEE Transactions on Services Computing}, 11(4):713--726, 2015.
		
		\bibitem{tang2014efficient}
		Zhuo Tang, Zhenzhen Cheng, Kenli Li, and Keqin Li.
		\newblock An efficient energy scheduling algorithm for workflow tasks in
		hybrids and dvfs-enabled cloud environment.
		\newblock In {\em 2014 Sixth International Symposium on Parallel Architectures,
			Algorithms and Programming}, pages 255--261. IEEE, 2014.
		
		\bibitem{guerout2013energy}
		Tom Gu{\'e}rout, Thierry Monteil, Georges Da~Costa, Rodrigo~Neves Calheiros,
		Rajkumar Buyya, and Mihai Alexandru.
		\newblock Energy-aware simulation with dvfs.
		\newblock {\em Simulation Modelling Practice and Theory}, 39:76--91, 2013.
		
		\bibitem{deelman2008cost}
		Ewa Deelman, Gurmeet Singh, Miron Livny, Bruce Berriman, and John Good.
		\newblock The cost of doing science on the cloud: the montage example.
		\newblock In {\em SC'08: Proceedings of the 2008 ACM/IEEE conference on
			Supercomputing}, pages 1--12. IEEE, 2008.
		
		\bibitem{jackson2010performance}
		Keith~R Jackson, Lavanya Ramakrishnan, Krishna Muriki, Shane Canon, Shreyas
		Cholia, John Shalf, Harvey~J Wasserman, and Nicholas~J Wright.
		\newblock Performance analysis of high performance computing applications on
		the amazon web services cloud.
		\newblock In {\em 2nd IEEE international conference on cloud computing
			technology and science}, pages 159--168. IEEE, 2010.
		
		\bibitem{madduri2015globus}
		Ravi Madduri, Kyle Chard, Ryan Chard, Lukasz Lacinski, Alex Rodriguez, Dinanath
		Sulakhe, David Kelly, Utpal Dave, and Ian Foster.
		\newblock The globus galaxies platform: delivering science gateways as a
		service.
		\newblock {\em Concurrency and Computation: Practice and Experience},
		27(16):4344--4360, 2015.
		
		\bibitem{ndamlabin2021dynamic}
		Jean~Etienne Ndamlabin~Mboula, Vivient~Corneille Kamla, and Cl{\'e}mentin
		Tayou~Djamegni.
		\newblock Dynamic provisioning with structure inspired selection and limitation
		of vms based cost-time efficient workflow scheduling in the cloud.
		\newblock {\em Cluster Computing}, 24(3):2697--2721, 2021.
		
		\bibitem{rodriguez2018scheduling}
		Maria~A Rodriguez and Rajkumar Buyya.
		\newblock Scheduling dynamic workloads in multi-tenant scientific workflow as a
		service platforms.
		\newblock {\em Future Generation Computer Systems}, 79:739--750, 2018.
		
		\bibitem{faragardi2019grp}
		Hamid~Reza Faragardi, Mohammad Reza~Saleh Sedghpour, Saber Fazliahmadi, Thomas
		Fahringer, and Nayereh Rasouli.
		\newblock Grp-heft: A budget-constrained resource provisioning scheme for
		workflow scheduling in iaas clouds.
		\newblock {\em IEEE Transactions on Parallel and Distributed Systems},
		31:1239--1254, 2019.
		
		\bibitem{singh2018novel}
		Vishakha Singh, Indrajeet Gupta, and Prasanta~K Jana.
		\newblock A novel cost-efficient approach for deadline-constrained workflow
		scheduling by dynamic provisioning of resources.
		\newblock {\em Future Generation Computer Systems}, 79:95--110, 2018.
		
		\bibitem{juve2013characterizing}
		Gideon Juve, Ann Chervenak, Ewa Deelman, Shishir Bharathi, Gaurang Mehta, and
		Karan Vahi.
		\newblock Characterizing and profiling scientific workflows.
		\newblock {\em Future Generation Computer Systems}, 29(3):682--692, 2013.
		
		\bibitem{topcuoglu2002performance}
		Haluk Topcuoglu, Salim Hariri, and Min-you Wu.
		\newblock Performance-effective and low-complexity task scheduling for
		heterogeneous computing.
		\newblock {\em IEEE transactions on parallel and distributed systems},
		13(3):260--274, 2002.
		
		\bibitem{ndamlabin2020cost}
		Jean~Etienne Ndamlabin~Mboula, Vivient~Corneille Kamla, and Cl{\'e}mentin
		Tayou~Djamegni.
		\newblock Cost-time trade-off efficient workflow scheduling in cloud.
		\newblock {\em Simulation Modelling Practice and Theory}, page 102107, 2020.
		
		\bibitem{kimura2006emprical}
		Hideaki Kimura, Mitsuhisa Sato, Yoshihiko Hotta, Taisuke Boku, and Daisuke
		Takahashi.
		\newblock Emprical study on reducing energy of parallel programs using slack
		reclamation by dvfs in a power-scalable high performance cluster.
		\newblock In {\em 2006 IEEE international conference on cluster computing},
		pages 1--10. IEEE, 2006.
		
		\bibitem{huang2012enhanced}
		Qingjia Huang, Sen Su, Jian Li, Peng Xu, Kai Shuang, and Xiao Huang.
		\newblock Enhanced energy-efficient scheduling for parallel applications in
		cloud.
		\newblock In {\em 2012 12th IEEE/ACM International Symposium on Cluster, Cloud
			and Grid Computing (ccgrid 2012)}, pages 781--786. IEEE, 2012.
		
		\bibitem{durillo2012moheft}
		Juan~J Durillo, Hamid~Mohammadi Fard, and Radu Prodan.
		\newblock Moheft: A multi-objective list-based method for workflow scheduling.
		\newblock In {\em 4th IEEE International Conference on Cloud Computing
			Technology and Science Proceedings}, pages 185--192. IEEE, 2012.
		
		\bibitem{singh2019energy}
		Vishakha Singh, Indrajeet Gupta, and Prasanta~K Jana.
		\newblock An energy efficient algorithm for workflow scheduling in iaas cloud.
		\newblock {\em Journal of Grid Computing}, pages 1--20, 2019.
		
		\bibitem{AmazonEC2ResModel}
		{Amazon EC2}.
		\newblock {Amazon EC2 Instance Types}.
		\newblock \url{https://aws.amazon.com/ec2/instance-types/}, 2019.
		\newblock Online; accessed 06 July 2019.
		
		\bibitem{AmazonEC2Pricing}
		{Amazon EC2}.
		\newblock {Amazon EC2 pricing}.
		\newblock \url{https://aws.amazon.com/ec2/pricing/on-demand/}, 2019.
		\newblock Online; accessed 06 July 2019.
		
		\bibitem{stavrinides2018impact}
		Georgios~L Stavrinides and Helen~D Karatza.
		\newblock The impact of workload variability on the energy efficiency of
		large-scale heterogeneous distributed systems.
		\newblock {\em Simulation Modelling Practice and Theory}, 89:135--143, 2018.
		
		\bibitem{wang2016hsip}
		Guan Wang, Yuxin Wang, Hui Liu, and He~Guo.
		\newblock Hsip: A novel task scheduling algorithm for heterogeneous computing.
		\newblock {\em Scientific Programming}, 2016, 2016.
		
		\bibitem{bharathi2008characterization}
		Shishir Bharathi, Ann Chervenak, Ewa Deelman, Gaurang Mehta, Mei-Hui Su, and
		Karan Vahi.
		\newblock Characterization of scientific workflows.
		\newblock In {\em 2008 third workshop on workflows in support of large-scale
			science}, pages 1--10. IEEE, 2008.
		
		\bibitem{calheiros2011cloudsim}
		Rodrigo~N Calheiros, Rajiv Ranjan, Anton Beloglazov, C{\'e}sar~AF De~Rose, and
		Rajkumar Buyya.
		\newblock Cloudsim: a toolkit for modeling and simulation of cloud computing
		environments and evaluation of resource provisioning algorithms.
		\newblock {\em Software: Practice and experience}, 41(1):23--50, 2011.
		
		\bibitem{mao2012performance}
		Ming Mao and Marty Humphrey.
		\newblock A performance study on the vm startup time in the cloud.
		\newblock In {\em 2012 IEEE Fifth International Conference on Cloud Computing},
		pages 423--430. IEEE, 2012.
		
	\end{thebibliography}

	\appendix
	\section{Further details on performance analysis}
	\label{appA}
	
	\subsection{Energy efficiency comparison through ANOVA with Tukey-Kramer post hoc tests}
	\label{appA:energy}

	\begin{landscape}	
		\begin{table*}[t]
			\setlength{\tabcolsep}{0.2em}
			
			\centering
			\caption{ANOVA test along with Tukey-Kramer pairwise tests comparing the Energy Consumption of SMWSO and SMWSH vs REEWS for MONTAGE 50, 100, 200, 500 and 10000} \label{tab_ANOVA_Tukey_pairwise_Test__Energy_Consumption_vs_REEWS_MONTAGE_ALL_Workload}
			
			\tiny
			
			
			\resizebox{\textwidth}{!}{\begin{tabular}{|l|c|c|c|c||  c  ||l|c|c|c|c|c|c||  c  ||l|c|c|c|c|}			
					\hline			
					\multicolumn{5}{|c||}{\textbf{\textit{ANOVA input summary}}}   
					&&  \multicolumn{7}{c||}{\textbf{\textit{ANOVA test result}}} 
					&& \multicolumn{5}{c|}{\textbf{\textit{Tukey-Kramer pairwise between the three algorithms}}} \\
					
					\hline
					\textit{\textbf{Group}} & \textit{\textbf{Count}} & \textit{\textbf{Sum}} & \textit{\textbf{Average}} & \textit{\textbf{Variance}} 
					&& 
					\textit{\textbf{Source of Variation}} & \textit{\textbf{SS}} & \textit{\textbf{df}} & \textit{\textbf{MS}} & \textit{\textbf{F stat}} & \textit{\textbf{P-value}} & \textit{\textbf{F crit}} 
					&&
					
					\textit{\textbf{Comparison}} & \textit{\textbf{Diff}} & \textit{\textbf{Abs. Diff}}  & \textit{\textbf{Q crit}} & \textit{\textbf{Is Sig?}} \\
					
					\hline	
					
					\textit{\textbf{SMWSO}} & 16	 & 72.16	 & 4.51	 & 8.41E-31  && \textit{\textbf{Between Groups}}  & 292.18	& 2	& 146.09	& 161.17	& \textbf{3.04E-21}	& 3.20
					&&\textit{\textbf{SMWSO vs SMWSH}}  & 2.25	& 2.25 & 0.81 & \textbf{YES}\\
					\hline	
					\textit{SMWSH} 			  & 16	 & 36.16	 & 2.26	 & 8.41E-31  && \textit{Within Groups} 			  & 40.79	& 45	& 0.91 &  &  & 
					&&\textit{\textbf{SMWSO vs REEWS}}  & -3.73	& 3.73 & 0.81 & \textbf{YES}\\
					\hline	
					\textit{REEWS} 			  & 16	 & 131.88	 & 8.24	 & 2.72   && \textit{Total} 			  		  & 332.97	& 47 &  &  &  & 
					&&\textit{\textbf{SMWSH vs REEWS}}  & -5.98	& 5.98 & 0.81 & \textbf{YES} \\
					\hline

					\multicolumn{19}{c}{\textbf{\textit{}}} \\
					
					\hline
					\multicolumn{19}{|c|}{\textbf{\textbf{\textit{Algorithm Ranking in terms of Energy saving}} : \textit{\textbf{\textit{SMWSO $\rightarrowtail 2$}} $|$ \textit{SMWSH $\rightarrowtail 1$}} $|$ \textbf{\textit{REEWS $\rightarrowtail 3$}}}} \\
					
					\hline
					
					\hline
					\multicolumn{19}{c}{(a) ANOVA test along with Tukey-Kramer pairwise tests comparing the Energy Consumption of SMWSO and SMWSH vs REEWS for \textbf{MONTAGE 50}} \\

					\multicolumn{19}{c}{\textbf{\textit{}}} \\
					\multicolumn{19}{c}{\textbf{\textit{}}} \\
					\hline			
					\multicolumn{5}{|c||}{\textbf{\textit{ANOVA input summary}}}   
					&&  \multicolumn{7}{c||}{\textbf{\textit{ANOVA test result}}} 
					&& \multicolumn{5}{c|}{\textbf{\textit{Tukey-Kramer pairwise between the three algorithms}}} \\
					
					\hline
					\textit{\textbf{Group}} & \textit{\textbf{Count}} & \textit{\textbf{Sum}} & \textit{\textbf{Average}} & \textit{\textbf{Variance}} 
					&& 
					\textit{\textbf{Source of Variation}} & \textit{\textbf{SS}} & \textit{\textbf{df}} & \textit{\textbf{MS}} & \textit{\textbf{F stat}} & \textit{\textbf{P-value}} & \textit{\textbf{F crit}} 
					&&
					
					\textit{\textbf{Comparison}} & \textit{\textbf{Diff}} & \textit{\textbf{Abs. Diff}}  & \textit{\textbf{Q crit}} & \textit{\textbf{Is Sig?}} \\
					
					\hline	
					
					\textit{\textbf{SMWSO}} & 16	 & 159.04	&9.94	&0  && \textit{\textbf{Between Groups}}  & 2563.40	&2	&1281.70	&93.84	&8,80E-17	&3.20
					&&\textit{\textbf{SMWSO vs SMWSH}}  & 4.43	&4.43 & 3.14 & \textbf{YES}\\
					
					\hline	
					
					\textit{SMWSH} 			  & 16	 & 88.16	&5.51	&8.41E-31  && \textit{Within Groups} 			  & 614.60	&45	&13.66	 &  &  & 
					&&\textit{\textbf{SMWSO vs REEWS}}  & -12.80	& 12.80 & 3.14 & \textbf{YES}\\
					
					\hline	
					
					\textit{REEWS} 			  & 16	 & 363.92	&22.745	&40.97 && \textit{Total} 			  		  & 3178	&47	 &  &  &  & 
					&&\textit{\textbf{SMWSH vs REEWS}}  & -17.23	& 17.23 & 3.14 & \textbf{YES} \\
					
					\hline

					\multicolumn{19}{c}{\textbf{\textit{}}} \\
					
					\hline
					\multicolumn{19}{|c|}{\textbf{\textbf{\textit{Algorithm Ranking in terms of Energy saving}} : \textit{\textbf{\textit{SMWSO $\rightarrowtail 2$}} $|$ \textit{SMWSH $\rightarrowtail 1$}} $|$ \textbf{\textit{REEWS $\rightarrowtail 3$}}}} \\
					
					\hline
					
					\hline
					\multicolumn{19}{c}{(b) ANOVA test along with Tukey-Kramer pairwise tests comparing the Energy Consumption of SMWSO and SMWSH vs REEWS for \textbf{MONTAGE 100}} \\

					\multicolumn{19}{c}{\textbf{\textit{}}} \\
					\multicolumn{19}{c}{\textbf{\textit{}}} \\
					\hline			
					\multicolumn{5}{|c||}{\textbf{\textit{ANOVA input summary}}}   
					&&  \multicolumn{7}{c||}{\textbf{\textit{ANOVA test result}}} 
					&& \multicolumn{5}{c|}{\textbf{\textit{Tukey-Kramer pairwise between the three algorithms}}} \\
					
					\hline
					\textit{\textbf{Group}} & \textit{\textbf{Count}} & \textit{\textbf{Sum}} & \textit{\textbf{Average}} & \textit{\textbf{Variance}} 
					&& 
					\textit{\textbf{Source of Variation}} & \textit{\textbf{SS}} & \textit{\textbf{df}} & \textit{\textbf{MS}} & \textit{\textbf{F stat}} & \textit{\textbf{P-value}} & \textit{\textbf{F crit}} 
					&&
					
					\textit{\textbf{Comparison}} & \textit{\textbf{Diff}} & \textit{\textbf{Abs. Diff}}  & \textit{\textbf{Q crit}} & \textit{\textbf{Is Sig?}} \\
					
					\hline	
					
					\textit{\textbf{SMWSO}} & 16	 & 373.12	&23.32	&1.35E-29  && \textit{\textbf{Between Groups}}  & 40379.09	&2	&121.83	&127.07	&6.90E-19	&3.20
					&&\textit{\textbf{SMWSO vs SMWSH}}  & 8.45	&8.45 & 10.94 & NO\\
					\hline	
					\textit{SMWSH} 			  & 16	 & 237.92	&14.87	&1.35E-29  && \textit{Within Groups} 			  & 7457.56	&45	&165.72 &  &  & 
					&&\textit{\textbf{SMWSO vs REEWS}}  & -56.86	&56.86 & 10.94 & \textbf{YES}\\
					\hline	
					\textit{REEWS} 			  & 16	 & 1282.96	&80.18	&497.17  && \textit{Total} 			  		  & 47836.65	&47 &  &  &  & 
					&&\textit{\textbf{SMWSH vs REEWS}}  & -65.31	&65.31 & 10.94 & \textbf{YES} \\
					\hline

					\multicolumn{19}{c}{\textbf{\textit{}}} \\
					
					\hline
					\multicolumn{19}{|c|}{\textbf{\textbf{\textit{Algorithm Ranking in terms of Energy saving}} : \textit{\textbf{\textit{SMWSO $\rightarrowtail 1$}} $|$ \textit{SMWSH $\rightarrowtail 1$}} $|$ \textbf{\textit{REEWS $\rightarrowtail 3$}}}} \\
					
					\hline
					
					\hline
					\multicolumn{19}{c}{(c) ANOVA test along with Tukey-Kramer pairwise tests comparing the Energy Consumption of SMWSO and SMWSH vs REEWS for \textbf{MONTAGE 200}} \\

					\multicolumn{19}{c}{\textbf{\textit{}}} \\
					\multicolumn{19}{c}{\textbf{\textit{}}} \\
					\hline			
					\multicolumn{5}{|c||}{\textbf{\textit{ANOVA input summary}}}   
					&&  \multicolumn{7}{c||}{\textbf{\textit{ANOVA test result}}} 
					&& \multicolumn{5}{c|}{\textbf{\textit{Tukey-Kramer pairwise between the three algorithms}}} \\
					
					\hline
					\textit{\textbf{Group}} & \textit{\textbf{Count}} & \textit{\textbf{Sum}} & \textit{\textbf{Average}} & \textit{\textbf{Variance}} 
					&& 
					\textit{\textbf{Source of Variation}} & \textit{\textbf{SS}} & \textit{\textbf{df}} & \textit{\textbf{MS}} & \textit{\textbf{F stat}} & \textit{\textbf{P-value}} & \textit{\textbf{F crit}} 
					&&
					
					\textit{\textbf{Comparison}} & \textit{\textbf{Diff}} & \textit{\textbf{Abs. Diff}}  & \textit{\textbf{Q crit}} & \textit{\textbf{Is Sig?}} \\
					
					\hline	
					
					\textit{\textbf{SMWSO}} & 16	 & 1456.32	&91.02	&0  && \textit{\textbf{Between Groups}}  & 2649602.10	&2	&1324801.05	&81.71	&1.05E-15	&3.20
					&&\textit{\textbf{SMWSO vs SMWSH}}  & 20.64	&20.64 & 108.20 & NO\\
					\hline	
					\textit{SMWSH} 			  & 16	 & 1126.02	&70.38	&2.5E-05 && \textit{Within Groups} 			  & 729598.73	&45	&16213.30 &  &  & 
					&&\textit{\textbf{SMWSO vs REEWS}}  & -487.75	& 487.75 & 108.20 & \textbf{YES}\\
					\hline	
					\textit{REEWS} 			  & 16	 & 9260.4	&578.77	&48639.91  && \textit{Total} 			  		  & 3379200.83	&47	 &  &  &  & 
					&&\textit{\textbf{SMWSH vs REEWS}}  & -508.40	&508.40 & 108.20 & \textbf{YES} \\
					\hline

					\multicolumn{19}{c}{\textbf{\textit{}}} \\
					
					\hline
					\multicolumn{19}{|c|}{\textbf{\textbf{\textit{Algorithm Ranking in terms of Energy saving}} : \textit{\textbf{\textit{SMWSO $\rightarrowtail 1$}} $|$ \textit{SMWSH $\rightarrowtail 1$}} $|$ \textbf{\textit{REEWS $\rightarrowtail 3$}}}} \\
					
					\hline
					
					\hline
					\multicolumn{19}{c}{(d) ANOVA test along with Tukey-Kramer pairwise tests comparing the Energy Consumption of SMWSO and SMWSH vs REEWS for \textbf{MONTAGE 500}} \\

					\multicolumn{19}{c}{\textbf{\textit{}}} \\
					\multicolumn{19}{c}{\textbf{\textit{}}} \\
					\hline			
					\multicolumn{5}{|c||}{\textbf{\textit{ANOVA input summary}}}   
					&&  \multicolumn{7}{c||}{\textbf{\textit{ANOVA test result}}} 
					&& \multicolumn{5}{c|}{\textbf{\textit{Tukey-Kramer pairwise between the three algorithms}}} \\
					
					\hline
					\textit{\textbf{Group}} & \textit{\textbf{Count}} & \textit{\textbf{Sum}} & \textit{\textbf{Average}} & \textit{\textbf{Variance}} 
					&& 
					\textit{\textbf{Source of Variation}} & \textit{\textbf{SS}} & \textit{\textbf{df}} & \textit{\textbf{MS}} & \textit{\textbf{F stat}} & \textit{\textbf{P-value}} & \textit{\textbf{F crit}} 
					&&
					
					\textit{\textbf{Comparison}} & \textit{\textbf{Diff}} & \textit{\textbf{Abs. Diff}}  & \textit{\textbf{Q crit}} & \textit{\textbf{Is Sig?}} \\
					
					\hline	
					
					\textit{\textbf{SMWSO}} & 16	 & 4160.32	&260.02	&0  && \textit{\textbf{Between Groups}}  & 30446883.33	&2	&15223441.66	&103.40	&1.50-17	&3.20
					&&\textit{\textbf{SMWSO vs SMWSH}}  & 35.93	&35.93 & 326.06 & NO\\
					\hline	
					\textit{SMWSH} 			  & 16	 & 3585.49	&224.09	&0.13  && \textit{Within Groups} 			  & 6625787.51	&45	&147239.72 &  &  & 
					&&\textit{\textbf{SMWSO vs REEWS}}  & -1671.24	&1671.24 & 326.06 & \textbf{YES}\\
					\hline	
					\textit{REEWS} 			  & 16	 & 30900.25	&1931.26	&441719.03   && \textit{Total} 			  		  & 37072670.84	&47	 &  &  &  & 
					&&\textit{\textbf{SMWSH vs REEWS}}  & -1707.17	&1707.17 & 326.06 & \textbf{YES} \\
					\hline

					\multicolumn{19}{c}{\textbf{\textit{}}} \\
					
					\hline
					\multicolumn{19}{|c|}{\textbf{\textbf{\textit{Algorithm Ranking in terms of Energy saving}} : \textit{\textbf{\textit{SMWSO $\rightarrowtail 1$}} $|$ \textit{SMWSH $\rightarrowtail 1$}} $|$ \textbf{\textit{REEWS $\rightarrowtail 3$}}}} \\
					
					\hline
					
					\hline
					\multicolumn{19}{c}{(e) ANOVA test along with Tukey-Kramer pairwise tests comparing the Energy Consumption of SMWSO and SMWSH vs REEWS for \textbf{MONTAGE 1000}} \\

					\multicolumn{19}{c}{\textbf{\textit{}}} \\
					\multicolumn{19}{c}{\textbf{\textit{}}} \\

			\end{tabular}}	
			

		\end{table*}
	\end{landscape}
\end{document}